\documentclass[english,american]{article}
\usepackage[T1]{fontenc}
\usepackage[latin9]{inputenc}
\usepackage{geometry}
\usepackage{textcomp}
\usepackage{amsbsy}
\usepackage{amssymb}
\usepackage{graphicx}
\usepackage{esint}
\usepackage{babel}
\begin{document}

\title{Identification of slow molecular order parameters for Markov model
construction}

\author{Guillermo Perez-Hernandez$^{1}$, Fabian Paul$^{1,4,+}$,\\
Toni Giorgino$^{2,+}$, Gianni de Fabritiis$^{3,\dagger}$, and Frank
Noé$^{1,*}$}
\maketitle
\begin{abstract}
A goal in the kinetic characterization of a macromolecular system
is the description of its slow relaxation processes, involving (i)
identification of the structural changes involved in these processes,
and (ii) estimation of the rates or timescales at which these slow
processes occur. Most of the approaches to this task, including Markov
models, Master-equation models, and kinetic network models, start
by discretizing the high-dimensional state space and then characterize
relaxation processes in terms of the eigenvectors and eigenvalues
of a discrete transition matrix. The practical success of such an
approach depends very much on the ability to finely discretize the
slow order parameters. How can this task be achieved in a high-dimensional
configuration space without relying on subjective guesses of the slow
order parameters? In this paper, we use the variational principle
of conformation dynamics to derive an optimal way of identifying the
``slow subspace'' of a large set of prior order parameters - either
generic internal coordinates (distances and dihedral angles), or a
user-defined set of parameters. It is shown that a method to identify
this slow subspace exists in statistics: the time-lagged independent
component analysis (TICA). Furthermore, optimal indicators---order
parameters indicating the progress of the slow transitions and thus
may serve as reaction coordinates---are readily identified. We demonstrate
that the slow subspace is well suited to construct accurate kinetic
models of two sets of molecular dynamics simulations, the 6-residue
fluorescent peptide MR121-GSGSW and the 30-residue natively disordered
peptide KID. The identified optimal indicators reveal the structural
changes associated with the slow processes of the molecular system
under analysis.
\end{abstract}
\vspace{0.5cm}

Adresses:

$1$ FU Berlin, Arnimallee 6, 14195 Berlin, Germany.

2 Institute of Biomedical Engineering (ISIB), National Research Council
of Italy (CNR), Corso Stati Uniti 4, I-35127 Padua, Italy.

3 GRIB, Barcelona Biomedical Research Park (PRBB), C/ Dr. Aiguader
88, 08003, Barcelona, Spain. 

4 Max-Planck-Institut für Colloids and Interfaces, Division Theory
and Bio-Systems, Science Park Potsdam-Golm, 14424 Potsdam, Germany

\vspace{0.5cm}

$+$ equal contribution

\vspace{0.5cm}

Corresponding authors:

$\dagger$ gianni.defabritiis@upf.edu 

{*} frank.noe@fu-berlin.de

\clearpage

\section{Introduction}

Conformational transitions between long-lived, or ``metastable''
states are essential to the function of biomolecules \cite{GansenSeidel_PNAS2009_Nucleosome,NeubauerSeidel_JACS2007_Rhodamine,Xie_PRL05_PowerLaw,EisenmesserKayKern_Nature2005,Santoso_PNAS2009_FretPolymerase,GebhardRief_PNAS10_AFMEnergyLandscapeProtein,WensleyClark_Nature09_FrustrationHelix}.
These rare transitions are ubiquitously found in biomolecular processes,
including folding \cite{JaegerEtAl_PNAS06,KobitskiNienhaus_NucleicAcidsRes07},
complex conformational rearrangements between native protein substates
\cite{FischerSmith_PNAS102_6873,NoeKrachtusSmithFischer_JCTC06_TransitionNetworks},
and ligand binding \cite{Nienhaus_Nature00}. Rare conformational
transitions can be explicitly traced by either single-molecule experiments
\cite{PirchiHaran_NatureComms11_SingleMoleculeFRET,StiglerRief_Science11_CalmodulinFoldingNetwork,ChungLuoisEaton_Science12_TransitionPathTimes,KobitskiNienhaus_NucleicAcidsRes07}
or by high-throughput molecular dynamics simulations, either realized
with few long trajectories \cite{Shaw_Science10_Anton,LindorffLarsenEtAl_Science11_AntonFolding}
or with many shorter trajectories \cite{VoelzPande_JACS10_NTL9,BowmanVoelzPande_JACS11_FiveHelixBundle-TripletQuenching,NoeSchuetteReichWeikl_PNAS09_TPT,BuchFabritiis_PNAS11_Binding,SadiqNoeFabritiis_PNAS12_HIV,Buch_Harvey_Giorgino_Anderson_DeFabritiis_2010}.
Molecular dynamics (MD) simulations are unique in their ability to
resolve the dynamics and all structural features of a biomolecule
simultaneously. When the sampling problem can be overcome and the
appropriateness of the force field parameters used is confirmed by
accompanying experimental evidence, MD simulations are amongst the
most powerful tools to investigate conformational transitions in biomolecules.

A current challenge with high-throughput MD simulations is to extract
meaningful information from vast trajectory data in an objective way.
To achieve this goal, the last few years have seen vast activity in
the development of computational methods that extract kinetic models
from the MD data. Kinetic models usually first partition the conformation
space into discrete states \cite{Wales,NoeFischer_CurrOpin08_TransitionNetworks,KarpenBrooks_Biochemistry93_Clustering,HubnerShakhnovich_PNAS96_EnsembleFolding,Weber_ImprovedPCCA,BucheteHummer_JPCB08,RaoCaflisch_JMB342_299,MuffCaflisch_Proteins07,deGrootDauraMarkGrubmuller_JMB301_299,SchultheisHirschbergerCarstensTavan_JCTC1_515,PanRoux_JCP08_MarkovModelPath}.
Subsequently, transition rates or probabilities can be estimated \cite{KrivovKarplus_PNAS101_14766,MuffCaflisch_Proteins07,NoeHorenkeSchutteSmith_JCP07_Metastability,ChoderaEtAl_JCP07,SwopePiteraSuits_JPCB108_6571,SchultheisHirschbergerCarstensTavan_JCTC1_515,PanRoux_JCP08_MarkovModelPath}.
The resulting models are often called transition networks \cite{RaoCaflisch_JMB342_299,NoeHorenkeSchutteSmith_JCP07_Metastability,HuangCaflisch_PlosCB11_SmallMoleculeUnbinding},
Diffusion maps \cite{CoifmanLafon_PNAS05_DiffusionMaps,RohrdanzClementi_JCP134_DiffMaps},
Master equation models \cite{SriramanKevrekidisHummer_JPCB109_6479,BucheteHummer_JPCB08},
Markov models \cite{Noe_JCP08_TSampling} or Markov state models \cite{SinghalPande_JCP123_204909,ChoderaEtAl_JCP07}
(MSM), where ``Markovianity'' means that the kinetics are modeled
by a memoryless jump process between states. 

The recent integration of classical statistical mechanics with modern
molecular kinetics highlights the crucial role of the eigenvectors
and eigenvalues of the Markov model transition matrix or Master equation
rate matrix. This is because they approximate the exact eigenfunctions
and eigenvalues of the propagator of the continuous dynamics \cite{SarichNoeSchuette_MMS09_MSMerror}.
The following eigenvalue equation is fundamental to conformation dynamics
\begin{equation}
\mathcal{P}\phi_{i}=\lambda_{i}\phi_{i}\label{eq_elementary-equation}
\end{equation}
Here, $\mathcal{P}$ is the transfer operator that propagates probability
densities of molecular configurations \cite{SchuetteFischerHuisingaDeuflhard_JCompPhys151_146,Noe_MMS12_VariationalPrinciple},
$\phi_{i}$ are its eigenfunctions, and $\lambda_{i}$ are the associated
eigenvalues. Equivalent expressions are obtained by expressing the
eigenfunctions in different weighted spaced, leading to the transfer
operator formulation \cite{SchuetteFischerHuisingaDeuflhard_JCompPhys151_146},
or the symmetrized propagator formulation \cite{BucheteHummer_JPCB08}.
Eq. (\ref{eq_elementary-equation}) is fundamental because when solving
it for the largest eigenvalues and associated eigenfunctions, all
stationary or kinetic quantities are defined by them. For example:
\begin{itemize}
\item $\mathcal{P}$ is guaranteed to have a unitary stationary eigenvalue
and the associated stationary distribution $\mu(\mathbf{x})=\phi_{1}(\mathbf{x})$;
The ensemble average of an observable $o$ can be calculated from
$o$ and $\mu$.
\item Experimentally measurable relaxation rates of the system can be computed
from the eigenvalues as $\kappa_{i}=-\tau^{-1}\ln\lambda_{i}$, or
the corresponding timescales as $t_{i}=\kappa_{i}^{-1}$.
\item The metastable states (often referred to as ``free energy basins''---although
we will avoid this term as it would imply the projection onto some
pre-defined coordinate set), can be computed from the sign structure
of the leading eigenfunctions \cite{SchuetteFischerHuisingaDeuflhard_JCompPhys151_146,DeuflhardWeber_PCCA}.
\item The structural transition associated to each relaxation timescale
is defined by the corresponding eigenfunction \cite{PrinzEtAl_JCP10_MSM1}
and corresponds to a transition between metastable sets. This fact
can be used to assign structural changes to experimentally measurable
timescales \cite{NoeEtAl_PNAS11_Fingerprints}.
\item Experimentally measurable correlation functions (e.g. fluorescence
correlation, intermediate scattering function in dynamic neutron or
X-ray scattering) can be computed as a sum of single-exponential relaxations
with timescales computed from $\lambda_{i}$ and amplitudes from the
$\phi_{i}$ and the experimental observable \cite{NoeEtAl_PNAS11_Fingerprints,KellerPrinzNoe_ChemPhysReview11,Buchner_BBA11_ProteinFoldingKinetics}.
\item From the largest $m$ Eigenvalues and their associated Eigenfunctions,
a rank-$m$ propagator can be assembled that can describe the dynamics
slower than timescale $t_{m}$ \cite{KubeWeber_JCP07_CoarseGraining}.
From this propagator, many properties can be calculated, such as transition
pathways between two sets of configurations \cite{MetznerSchuetteVandenEijnden_TPT,BerezhkovskiiHummerSzabo_JCP09_Flux,NoeSchuetteReichWeikl_PNAS09_TPT}.
\end{itemize}
The approximation error of all of the above quantities can be cast
in terms of the approximation error of the eigenvalues and eigenfunctions
\cite{SarichNoeSchuette_MMS09_MSMerror,DjurdjevacSarichSchuette_MMS10_EigenvalueError,PrinzEtAl_JCP10_MSM1,PrinzChoderaNoe_PRL11_RateTheory}.
Vice versa, all of the above quantities are easily and precisely computable
when the eigenvalues and eigenfunctions of $\mathcal{P}$ have been
approximated with high precision. Consequently, any modeling method
that attempts to compute the above quantities must aim at approximating
the eigenvalues and eigenfunctions of $\mathcal{P}$ - either explicitly
or implicitly. 

Markov models and most of the other aforementioned kinetic models
require a discretization of configuration space to be made. This is
typically done by choosing representative configurations by some data
clustering method, and then partitioning the configuration space by
a Voronoi tesselation. In contrast to other fields of data analysis,
the purpose of clusters is not a classification of configurations,
but rather a sufficiently fine discretization of configuration space
such that the eigenfunctions can be well approximated in terms of
step functions on the Voronoi cells \cite{PrinzEtAl_JCP10_MSM1}.
In order to achieve this, the metric must be chosen such that a fine
partition of the relevant ``slow'' order parameters, i.e. those
which are good indicators of the slow eigenfunctions $\phi_{i}$. 

How can the slow order parameters be identified without already having
a high-precision Markov model? It was early noted that a priori order
parameters, such as the root mean square distance (RMSD) to a single
reference structure, the radius of gyration, or pre-selected distances
or angles are often not good indicators of the slow eigenfunctions,
and thus bear the danger of disguising the slow kinetics \cite{KrivovKarplus_PNAS101_14766,NoeFischer_CurrOpin08_TransitionNetworks,MuffCaflisch_Proteins07}.
In order to avoid this, Markov model construction has focused in the
last years focused on the other extreme - using general metrics that
are capable of describing every sort of configurational change. Most
notable is the minimal RMSD metric, which assigns to each pair of
configurations their minimal Euclidean distance subject to rigid-body
translation and rotation \cite{Kabsch::76}. Minimal RMSD has been
used successfully in many examples, especially protein folding (see
\cite{Bowman_JCP09_Villin,PrinzEtAl_JCP10_MSM1} and references therein).
Recent applications include folding of MR121-GSGS-W peptide\cite{PrinzEtAl_JCP10_MSM1},
folding of FiP35 WW domain, GTT, NTL9, and protein G \cite{Beauchamp::12}
and discovery of cryptic allosteric sites in $\beta$-lactamase, interleukin-2,
and RNase H \cite{Bowman::12}. However, minimal RMSD tends to fail
in situation where the largest-amplitude motions are not the slowest
(an example of this is the natively disordered KID peptide analyzed
below). Principal component analysis (PCA) is a frequently-used method
to reduce the dimension of an order parameter space by projecting
it on its linear subspace of the largest-amplitude motions \cite{Amadei_Proteins17_412}.
PCA has also been used successfully in Markov model construction \cite{NoeSchuetteReichWeikl_PNAS09_TPT,NoeEtAl_PNAS11_Fingerprints},
however is suffers from the similar limitations like minimal RMSD,
as there is no general guarantee that large-amplitude motions are
associated with slow transitions. 

It is an important challenge to find a metric that provides a good
indicator of the slow processes, such that a good approximation of
the eigenfunctions $\phi_{i}$ is feasible with a moderate number
of clusters. The aim of this paper is to identify such a method. To
be more precise, let $r_{1},...,r_{d}\in\mathbb{R}$ be a possibly
large set of $d$ order parameters of a molecular system that are
\emph{a priori} specified by the user. Typical examples of order parameter
include intramolecular distances and torsion angles. However, complex
order parameters like the instantaneous dipole moment of a molecule,
or an experimentally measurable quantity such as a FRET efficiency
may also be included. Given this set of order parameters, we aim to
\begin{enumerate}
\item Find the linear combination of order parameters that optimally approximates
the dominant eigenvalues and eigenfunctions, such that a high-precision
Markov model can be built in these order parameters with direct clustering.
\item Identify the $m$ order parameters that are best and least redundant
indicators for the $m$ dominant eigenfunctions, thus providing the
user a direct physical interpretation which structural changes are
associated with the slowest relaxation timescales (Feature selection).
\end{enumerate}
Here we use the variational principle of conformation dynamics \cite{Noe_MMS12_VariationalPrinciple}
to derive an optimal solution for problem 1, and show that an existing
extension to PCA solves this problem: time-lagged independent component
analysis (TICA) combines information from the covariance matrix and
a time-lagged covariance matrix of the data \cite{Molgedey::94}.
See \cite{HyvaerinenKarhunenOja_ICA_Book} for a detailed description
of the method. TICA has recently been applied in the analysis of MD
data. Naritomi and Fuchigami\cite{Naritomi::11} used TICA to investigate
domain motion of the LAO protein and compared it to PCA. Mitsutake
\emph{et al} \cite{Mitsutake::11} used relaxation mode analysis,
a related technique, to analyze the dynamics of Met-enkephalin. Both
studies showed that the slow modes were not necessarily associated
with large amplitudes, and time-lagged mode analyses were thus better
suited to detect them than PCA. Here, we demonstrate the usefulness
of TICA coordinates for constructing Markov models for two rather
different molecular processes: (i) the conformational dynamics of
the small fluorescent peptide MR121-GSGSW, for which good Markov models
can be built using a variety of methods, and of the natively unstructured
30-residue peptide KID, modeled through a large ensemble of explicit-solvent
molecular dynamics (MD) simulations.

We also propose a way to approach problem 2, identifying the optimal
indicators of the slowest processes. These indicators inform the user
of the structural process that is governing the slow relaxations of
the macromolecule. Optimal indicators help in understanding what comprises
the slow kinetics, and dramatically the user time to ``search''
for a structural character of the slow processes from a Markov model.

\section{Theory}

We summarize the variational principle of conformation dynamics, stating
that the true eigenfunctions are best approximated by a Markov model,
when the estimated timescales $\hat{t}_{i}$ are maximized. We derive
a way to optimally approximate the true eigenfunctions in terms of
a linear combination of the original order parameters. We then show
that this method is identical to the time-lagged independent component
analysis (TICA) that is an established method in statistics. The TICA
problem can be easily solved by subsequently solving two simple Eigenvalue
problems.

\subsection{Exact dynamics in full configuration space}

We start by providing an expression for the propagator of exact continuous
molecular dynamics, and show that in order to approximate its long-time
behavior, its largest eigenvalues and associated eigenfunctions must
be well approximated. 

We use $\mathbf{x}_{t}$ to denote the full molecular configuration
at time $t$ (if velocities are available, $\mathbf{x}_{t}$ denotes
a point in full phase space) in state or phase space $\Omega$. We
assume that the molecular dynamics implementation is Markovian in
$\Omega$ (i.e. the time step to $\mathbf{x}_{t+\tau}$ is computed
based on the current value of $\mathbf{x}_{t}$ only), and gives rise
to a unique stationary density $\mu(\mathbf{x})$, usually the Boltzmann
density:
\[
\mu(\mathbf{x})=Z^{-1}\mathrm{e}^{-\beta H(\mathbf{x})}.
\]
where $H$ is the Hamiltonian, $Z$ is the partition function, and
$\beta=(k_{B}T)^{-1}$ is the inverse temperature. We also assume
that the dynamics are statistically reversible, i.e. that the molecular
system is simulated in thermal equilibrium. Let us denote a probability
density of molecular configurations as $\rho_{t}$, and let us subsume
the action of the molecular dynamics implementation into the propagator
$\mathcal{P}(\tau)$. The propagator describes the probability that
a trajectory that is at configuration $\mathbf{x}_{t}$ at time $t$
will be found at a configuration $\mathbf{x}_{t+\tau}$ a time $\tau$
later. In an ensemble view, the propagator takes a probability density
of configurations, $\rho_{t}$ and predicts the probability density
of configurations at later time, $\rho_{t+\tau}$:
\[
\rho_{t+\tau}=\mathcal{P}(\tau)\rho_{t}
\]

We can write the propagator, by expanding it in terms of its eigenvalues:
\[
\lambda_{i}(\tau)=\mathrm{e}^{-\frac{\tau}{t_{i}}}
\]
and its eigenfunctions $\phi_{i}$, as:
\begin{eqnarray}
\rho_{t+\tau}(\mathbf{y})=\mathcal{P}(\tau)\rho_{t}(\mathbf{x}) & = & \sum_{i=1}^{\infty}\mathrm{e}^{-\frac{\tau}{t_{i}}}\langle\psi_{i},\rho_{t}\rangle\phi_{i}.\label{eq_propagator-spectral}
\end{eqnarray}
where the eigenfunctions $\phi_{i}(\mathbf{x})$ take the role of
basis functions with which probability densities $\rho$ can be constructed.
The first eigenvalue is $\lambda_{1}=1$ and the remaining eigenvalues
have a norm strictly smaller than $1$. Thus, the first timescale
is $t_{1}=\infty$ and corresponds to the stationary distribution,
while all other timescales $t_{i}$ are finite relaxation timescales.
$\psi_{i}(\mathbf{x})=\mu^{-1}(\mathbf{x})\phi_{i}(\mathbf{x})$ are
the eigenfunctions weighted by the stationary density. Eq. (\ref{eq_propagator-spectral})
has a straightforward physical interpretation: the scalar product
$\langle\psi_{i},\rho_{t}\rangle$ measures the overlap of the starting
density $\rho_{t}$ with the $i$th eigenfunction and thus determines
the amplitude by which this eigenfunction contributes to the dynamics.
At any time $\tau$, the new probability density $\rho_{t+\tau}$
is composed of a set of basis functions $\phi_{i}$. With increasing
time, the contributions of all basis functions $\phi_{i}$ with $i>1$
vanish exponentially with a timescale given by $t_{i}$. After infinite
time $\tau\rightarrow\infty$, only the first term with $t_{1}=\infty$
(and hence $\exp(-\tau/t_{i})=1$) is left, and the stationary density
is reached: $\lim_{\tau\rightarrow\infty}\mathcal{P}(\tau)\rho_{t}=\phi_{1}=\mu$.
Stationarity implies that $\mu$ will not be changed under the action
of the propagator:
\[
\mathcal{P}(\tau)\mu=\mu.
\]

Suppose we are interested in slow timescales $\tau\gg t_{m+1}$. At
such large times, the dynamics is governed by the $m$ largest timescales
$t_{i}$ and eigenfunctions of the propagator:
\[
\rho_{t+\tau}=\mathcal{P}(\tau)\rho_{t}\approx\sum_{i=1}^{m}\mathrm{e}^{-\frac{\tau}{t_{i}}}\langle\psi_{i},\rho_{t}\rangle\phi_{i}.
\]
All kinetic properties at this timescale and all stationary properties
can be accurately computed when the dominant $m$ eigenvalues and
eigenfunctions are approximated. This is our goal.

\subsection{Approximation of slowest timescales and the related eigenfunctions}

We can make a few general statements on how to approximate the true
timescales $t_{i}$ and eigenfunctions. These general properties can
be used to derive a general method that achieves the aim of this paper:
the identification of the slowest order parameters in a molecule.
Since $\phi_{i}$ and $\psi_{i}$ are interchangeable using the weights
$\mu$, the approximation problem can be described using either kind
of eigenfunction. Subsequently we will always refer to the problem
of approximating the weighted eigenfunctions $\psi_{i}$. 

Consider some function of the molecular configuration, $f(\mathbf{x})$.
From Eq. (\ref{eq_propagator-spectral}) we can express the time-autocorrelation
function of $f$ as a function of $\tau$ as: 
\begin{equation}
\langle f(\mathbf{x}_{t})f(\mathbf{x}_{t+\tau})\rangle_{t}=\sum_{i=1}^{\infty}\mathrm{e}^{-\frac{\tau}{t_{i}}}\langle\phi_{i},f\rangle^{2}\label{eq_autocorrelation-spectral}
\end{equation}
Suppose we would know the true eigenfunction $\psi_{i}(\mathbf{x})$.
It is now easy to show \cite{Noe_MMS12_VariationalPrinciple} that
the time-autocorrelation function of $\psi_{i}(\mathbf{x})$ yields
the exact $i$th eigenvalue, and thus permits to recover the exact
$i$th timescale:
\begin{eqnarray*}
\hat{\lambda}_{i}(\tau) & = & \langle\psi_{i}(\mathbf{x}_{t})\psi_{i}(\mathbf{x}_{t+\tau})\rangle=\mathrm{e}^{-\frac{\tau}{t_{i}}}\\
\hat{t}_{i} & = & -\frac{\tau}{\ln|\hat{\lambda}_{i}(\tau)|}=t_{i}.
\end{eqnarray*}
However, in reality we will not know the exact eigenfunction $\psi_{i}$.
Suppose that we would \emph{guess} a model function $\hat{\psi}_{2}$
that is supposed to be similar to $\psi_{2}$. When we make sure that
$\hat{\psi}_{2}$ is appropriately normalized, the variational principle
of conformation dynamics \cite{Noe_MMS12_VariationalPrinciple} shows
that the time-autocorrelation function of $\hat{\psi}_{2}$ approximates
the true eigenvalue, and the true timescale from below:
\begin{eqnarray}
\langle\hat{\psi}_{2}(\mathbf{x}_{t})\hat{\psi}_{2}(\mathbf{x}_{t+\tau})\rangle & \le & \mathrm{e}^{-\frac{\tau}{t_{2}}}\nonumber \\
\hat{t}_{2} & \le & t_{2}\label{eq_t2-approximation}
\end{eqnarray}
where equality only holds for $\hat{\psi}_{2}=\psi_{2}$. Thus, we
have a recipe for finding an optimal approximation to the second timescale
and its associated eigenfunction: We must seek a function $\hat{\psi}_{2}$
that has the maximum timescale $\hat{t}_{2}$. 

Similar inequalities can be shown for the other eigenvalues and timescales
$t_{3},...,t_{m}$. We can show that if one proposes a model function
$\hat{\psi}_{i}$ that is orthogonal to the eigenfunctions $1$ through
$i-1$, we also have:
\begin{equation}
\hat{t}_{i}\le t_{i}.\label{eq_ti-approximation}
\end{equation}
This variational principle of conformation dynamics is analogous to
the variational principle in quantum mechanics.

\subsection{Best approximation of the eigenfunctions}

What is the relation of the variational principle above to Markov
models? Since the eigenfunctions $\psi_{i}$ are initially unknown
and difficult to guess, it is reasonable to approximate them by functions
$\hat{\psi}_{i}$ that are assembled from a linear combination of
basis functions
\begin{equation}
\hat{\psi}_{i}(\mathbf{x})=\sum_{k=1}^{n}b_{ik}\chi_{k}(\mathbf{x})\label{eq_linear-combination}
\end{equation}
which must be defined a priori, and the optimization problem then
consists of finding the optimal parameters $b_{ik}$ that we will
denote by vectors $\mathbf{b}_{i}\in\mathbb{R}^{n}$, where we have
chosen the dimension of the basis set, $n$, to be equal to the number
of basis functions. The Ritz method \cite{Ritz_JReineAngewMathe09_Variationsprobleme}
provides the optimal set of coefficients for an orthonormal basis
set. Formally, if we define the covariance matrix between Ansatz functions
as:
\[
c_{ij}^{\chi}(\tau)=\mathbb{E}_{t}[\chi_{i}(\mathbf{x}_{t})\chi_{j}(\mathbf{x}_{t+\tau})]
\]
And we require that the basis functions are orthogonal---which is
equivalent to them being uncorrelated at lag time 0:
\begin{equation}
\langle\chi_{i},\chi_{j}\rangle_{\mu}=\mathbb{E}_{t}[\chi_{i}(\mathbf{x}_{t})\chi_{j}(\mathbf{x}_{t})]=c_{ij}^{\chi}(0)=\delta_{ij}\label{eq_covariance-matrix}
\end{equation}
then the optimal set of coefficients is then given by the eigenvectors
$\mathbf{b}_{i}$ of the following Eigenvalue problem:
\begin{equation}
\mathbf{C}^{\chi}(\tau)\mathbf{b}_{i}=\mathbf{b}_{i}\hat{\lambda}_{i}(\tau)\label{eq_Ritz}
\end{equation}
Let us now consider the more general case that the Ansatz functions
are not orthonormal, i.e. $\langle\chi_{i},\chi_{j}\rangle_{\mu}\ne\delta_{ij}$.
In this situation we must first orthonormalized the basis coordinates
before. This is done via a generalization to Eq. (\ref{eq_Ritz}).
For a non-orthonormal basis set, the optimal approximation to the
true eigenvalues and eigenfunctions is obtained by solving the generalized
eigenvalue problem: 
\begin{equation}
\mathbf{C}^{\chi}(\tau)\mathbf{b}_{i}=\mathbf{C}^{\chi}(0)\mathbf{b}_{i}\hat{\lambda}_{i}(\tau)\label{eq_Roothaan-Hall}
\end{equation}
One may formally rewrite Eq. (\ref{eq_Roothaan-Hall}) as $\mathbf{D}(\tau)\mathbf{b}_{i}=\hat{\lambda}_{i}\mathbf{b}_{i}$
where $\mathbf{D}(\tau)=(\mathbf{C}^{\chi}(0))^{-1}\mathbf{C}^{\chi}(\tau)$
is an orthonormal basis set. Numerically, the matrix inversion of
$\mathbf{C}^{\chi}(0)$ is often poorly conditioned and should therefore
be avoided. The results (\ref{eq_Ritz}) and (\ref{eq_Roothaan-Hall})
are well known from variational calculus. The appendix contains an
illustrative derivation of Eq. (\ref{eq_Roothaan-Hall}) relevant
to the special choice of basis set used in this paper.

\subsection{Optimal linear combination of input order parameters}

\label{sub_theory_optimal-linear-combination}

Based on the above results we can now formulate a method to find a
linear combination of molecular order parameters $\mathbf{r}=(r_{1}(\mathbf{x}),...,r_{d}(\mathbf{x}))$
that best resolves the slow relaxation processes. This is done by
finding the optimal coefficients for Eq. (\ref{eq_linear-combination}).
For this, we define the Basis function $\chi_{i}$ to be identical
to the mean-free coordinate $r_{i}(\mathbf{x})$ (if the original
order parameters $r_{i}'(\mathbf{x})$ are not mean-free, then we
simply subtract the mean: $r_{i}(\mathbf{x})=r_{i}'(\mathbf{x})-\langle r_{i}(\mathbf{x})\rangle$):
\begin{equation}
\chi_{i}(\mathbf{x})=r_{i}(\mathbf{x})\label{eq_normalized-coordinates}
\end{equation}
Thus, our basis set has $n=d$ dimensions. Now let us compute the
correlation matrix of normalized order parameters as:
\[
c_{ij}^{r}(\tau)=\mathbb{E}_{t}[r_{i}(\mathbf{x}_{t})r_{j}(\mathbf{x}_{t+\tau})]=c_{ij}^{\chi}(\tau)
\]
Then solving Eq. (\ref{eq_Roothaan-Hall}) with the correlation matrix
for lag times 0 and $\tau$ will provide us with the linear combination
of input order parameters that optimally approximates the exact propagator
eigenfunctions. See Appendix for a sketch of the usual derivation
of Eq. (\ref{eq_Roothaan-Hall}) for the case of TICA. It turns out
that Eq. (\ref{eq_Roothaan-Hall}) with the choice of coordinates
(\ref{eq_normalized-coordinates}) is known as the time-lagged independent
component analysis (TICA) in statistics \cite{Molgedey::94,Naritomi::11}.
A robust algorithm to solve Eq. (\ref{eq_Roothaan-Hall}) is known
as AMUSE algorithm \cite{Tong::90} and will be given below.

The Eigenfunction approximations via Eq. (\ref{eq_linear-combination})
using the coefficients $\mathbf{b}_{i}$ are the optimal approximation
to the true eigenfunctions and will give an optimal approximation
of the timescales. As a result of the variational principle, $\hat{\lambda}_{2}(\tau)\le\lambda_{2}(\tau)$
and
\[
\hat{t}_{2}(\tau)=-\frac{\tau}{\ln\hat{\lambda}_{2}(\tau)}\le t_{2}
\]
according to \ref{eq_t2-approximation}. Since the true eigenfunctions
are generally nonlinear functions of the original order parameters,
and the basis set used in Eq. (\ref{eq_normalized-coordinates}) is
linear in the original order parameters, it cannot be expected that
$\boldsymbol{\psi}_{2}\approx\hat{\boldsymbol{\psi}}_{2}$ is true,
and therefore the variational principle can at this point not be extended
to further timescales than $t_{2}$. In other words, the TICA timescales
$\hat{t}_{3},...,\hat{t}_{m}$ may be both under- or overestimated.

\subsection{Markov models and implied timescales}

We do not intend to use the TICA timescales directly, but rather use
the TICA subspace in order to construct a Markov model by finely discretizing
this space. What can be said about the timescales of the resulting
Markov model? We can use the variational principle summarized above
to bound the timescales of a Markov model. Classical Markov models
operate by assigning a configuration $\mathbf{x}$ uniquely to one
of the geometric clusters used to construct them. It can be shown
\cite{SarichNoeSchuette_MMS09_MSMerror} that this operation is equivalent
to use the basis functions
\[
\chi_{i}(\mathbf{x})=\frac{\mathbf{1}_{i}(\mathbf{x})}{\sqrt{\pi_{i}}},
\]
i.e. each basis function $i$ is a step function with has a constant
value on the configurations belonging to the $i$th cluster and is
zero elsewhere. This basis is an orthonormal basis set: $\langle\chi_{i},\chi_{j}\rangle_{\mu}=\pi_{i}^{-1}\int_{\mathbf{x}\in S_{i}}\mu(\mathbf{x})\: d\mathbf{x}=\delta_{ij}$.
Thus, the direct Ritz method applies and as shown in \cite{Noe_MMS12_VariationalPrinciple},
Eq. (\ref{eq_Ritz}) becomes:
\begin{equation}
\mathbf{T}(\tau)\mathbf{r}_{i}=\mathbf{r}_{i}\tilde{\lambda}_{i}(\tau)\label{eq_MSM-Eigenvectors}
\end{equation}
where $\mathbf{T}(\tau)$ is the Markov model transition matrix and
$\mathbf{R}=[\mathbf{r}_{1},...,\mathbf{r}_{n}]$ are its right eigenvectors.
To relate Eq. (\ref{eq_Ritz}) and (\ref{eq_MSM-Eigenvectors}) we
have used the definition $\mathbf{C}^{\chi}(\tau)=\sqrt{\frac{\pi_{i}}{\pi_{j}}}T_{ij}(\tau)$,
i.e. the covariance matrix between Ansatz functions $\chi$ is the
symmetrized transition matrix as given in \cite{BucheteHummer_JPCB08}.

Thus, a Markov model is the Ritz method for the choice of a step-function
basis on the clusters used to build it, and thus gives an optimal
step-function approximation to the eigenfunctions and maximal eigenvalues
amongst all choices of functions that can be supported by the clustering.
It follows from Eq. (\ref{eq_t2-approximation}) that at least the
second timescale will then be underestimated. When the Markov model
is sufficiently good in approximating the slowest processes, all of
the first $m$ timescales will be underestimated as given by Eq. (\ref{eq_ti-approximation}).
It was shown \cite{DjurdjevacSarichSchuette_MMS10_EigenvalueError}
that this estimation error becomes smaller when $\tau$ is increased.
Prinz et al \cite{PrinzChoderaNoe_PRL11_RateTheory} showed that it
decreases with $\tau^{-1}$. As a result, when plotting the estimated
timescales $\hat{t}_{i}(\tau)$ as a function of $\tau$ one obtains
the well-known implied timescale plots shown in Figs. (\ref{fig_1})
and (\ref{fig_3}), where the estimated timescales $\hat{t}_{i}(\tau)$
slowly converge to the true timescale when $\tau$ is increased.

\begin{figure}
\noindent \begin{centering}
\includegraphics[width=0.8\columnwidth]{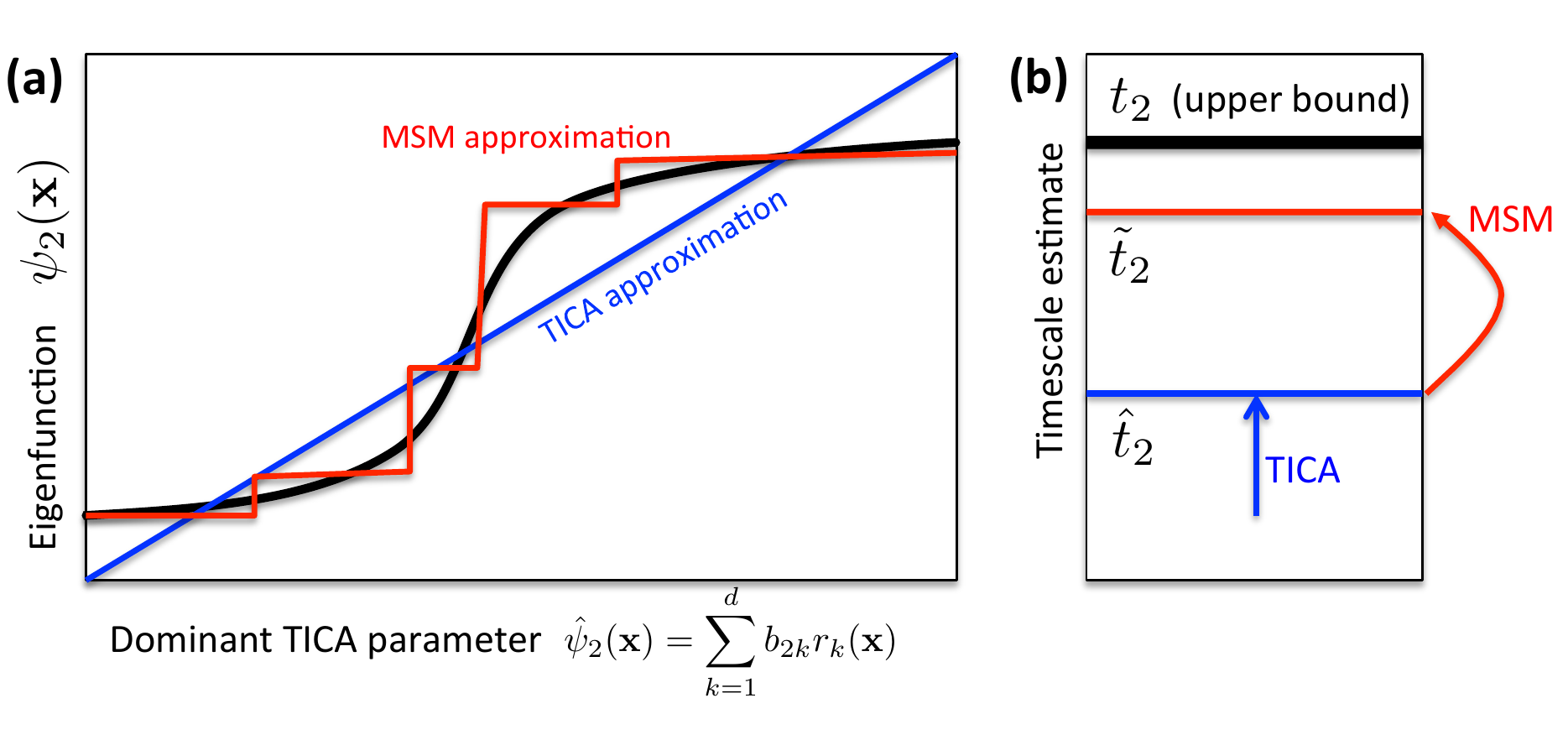}
\par\end{centering}

\caption{\label{fig_approximation-scheme}Scheme illustrating different approximations
to the dominant eigenfunction $\psi_{2}$ of the Molecular dynamics
propagator, and the associated approximations to the slowest relaxation
timescale $t_{2}$. TICA (blue) approximates the eigenfunction $\psi_{2}$
(black) as a linear combination of molecular observables and the TICA
timescale $\hat{t}_{2}$ associated to the TICA eigenvalue $\hat{\lambda}_{2}$
underestimates (usually strongly) the true timescale $t_{2}$. The
estimate is then improved by building a Markov model in TICA space
which approximates the eigenfunction $\psi_{2}$ by a step function
(red) that is constant on the Markov model clusters. The corresponding
Markov model estimate of the relaxation timescale, $\tilde{t}_{2}$,
is thus typically larger than the TICA timescale $\hat{t}_{2}$ and
a better estimate of the true timescale $t_{2}$.}
\end{figure}

We have now seen that both the TICA eigenvalue \foreignlanguage{english}{$\hat{\lambda}_{2}$}
and the corresponding timescale $\hat{t}_{2}$ are underestimated,
as well as the Markov model eigenvalue $\tilde{\lambda}_{2}$ and
the corresponding timescale $\tilde{t}_{2}$. Unfortunately, we cannot
make a rigorous statement of how $\hat{t}_{2}$ and $\tilde{t}_{2}$
are related to each other. However, we can make the \emph{ad hoc}
statement that we intend to cluster the dominant TICA subspace ``sufficiently
fine''. Thereby the Markov model step functions of the dominant TICA
component allows the nonlinear eigenfunction $\psi_{2}(\mathbf{x})$
to be approximated better than by the linear combination of order
parameters (\ref{eq_normalized-coordinates}) directly. For example,
it is typical that the eigenfunction $\psi_{2}(\mathbf{x})$ stays
almost constant over a large part of configuration space and then
changes abruptly to a different level in the transition state \cite{SchuetteFischerHuisingaDeuflhard_JCompPhys151_146,PrinzEtAl_JCP10_MSM1}.
Such a behavior can be much better described by a step function than
by a linear fit. Therefore, we shall here assume that the estimates
of the dominant timescale as $\hat{t}_{2}<\tilde{t}_{2}<t_{2}$: The
dominant TICA timescale $\hat{t}_{2}$ is a lower bound to the true
timescale $t_{2}$, but typically a poor lower bound. The Markov model
timescale $\tilde{t}_{2}$ is typically larger, and thus a better
estimate of the true timescale $t_{2}$. This concept is illustrated
in Fig. \ref{fig_approximation-scheme}.

\section{Methods}

Having identified the ``slow'' linear combinations of input order
parameters, the hope is that clustering in a low-dimensional linear
subspace will provide a useful clustering metric for the accurate
and efficient construction of Markov models with a moderate number
of clusters. Here, we compare the performance of different cluster
metrics which are briefly described in the present section. 

There are many software packages available for performing data clustering.
For clustering and Markov model construction of molecular dynamics
data, the packages EMMA \cite{SenneSchuetteNoe_JCTC12_EMMA1.2}, MSMbuilder
\cite{BeauchampEtAl_MSMbuilder2}, Wordom \cite{SeeberCaflisch_JCC11_Wordom}
and METAGUI \cite{Biarnes_Pietrucci_Marinelli_Laio_2012} are currently
available. Here, we use the EMMA package.

\subsection{Clustering methods and partitioning of state space}

Clustering methods for discretizing MD trajectory data can be divided
into two categories:
\begin{enumerate}
\item explicit coordinate methods that treat molecular coordinates as elements
of an explicit vector space. The MD data is projected into the chosen
coordinate set and then clustered by some distance metric (e.g. Euclidean
distance) in that space.
\item pure metric methods that have no explicit vector space available,
but rather a metric that measures the distance between pairs of molecular
conformations. The clustering algorithm groups only existing configurations
\emph{via} this distance metric.
\end{enumerate}
Coordinates used in (1) include Cartesian coordinates (provided there
is a meaningful coordinate origin, for example defined by the largest
molecule or domain in the system). For example, in Refs. \cite{HeldEtAl_BiophysJ10_AssociationTPT,BuchFabritiis_PNAS11_Binding},
the binding of a small ligand was analyzed by using the three-dimensional
positions of the ligand with respect to the protein. Other frequently
used coordinate sets are intramolecular coordinates such as dihedral
angles \cite{NoeHorenkeSchutteSmith_JCP07_Metastability} or inter-atomic
distances. Alternatively to directly clustering the primary set of
coordinates, coordinate transforms can be applied to preprocess the
data, with the hope of identifying sub-spaces in which the clustering
will be more informative. Below we will discuss the transformations
PCA and TICA in detail. 

A metric often used in (2) is the normalized Euclidean metric after
rigid-body translation and rotation has been removed, in short the
minimal RMSD (or least RMSD) metric \cite{Kabsch::76}. Minimal RMSD
is an established metric for the analysis of MD trajectories and MSM
construction and used here as a reference.

Discretization of trajectory data is performed using clustering algorithms.
In a first stage the trajectory is explored and $n$ representative
points of the coordinate space are selected as \emph{cluster centers}
with a clustering method. Various algorithms have been proposed in
the literature, including $k$-means \cite{Lloyd_IEEE82_kMeans},
$k$-centers \cite{DasguptaLong_JComputSystSci05_kCenters}, $k$-medoids
\cite{KaufmanRousseeuw_kMedoids}, regular spatial clustering\cite{SenneSchuetteNoe_JCTC12_EMMA1.2},
regular temporal clustering\cite{SenneSchuetteNoe_JCTC12_EMMA1.2}
and Ward clustering\cite{Beauchamp::12}. The $k$-means algorithm
requires the coordinate space to be a vector space (in order to compute
the mean) whereas the other aforementioned algorithms only require
a metric. 

Subsequent to the identification of cluster centers, the state space
is partitioned by assigning each trajectory frame to its closest cluster
center according to the same metric used for clustering. The discretization
obtained this way is a Voronoi tessellation of the observed coordinate
space. Voronoi cells form a complete partition of the conformation
space.

\subsection{Principal component analysis (PCA)}

PCA is a linear transform that transforms coordinates in such a way
that their instantaneous correlations vanishes. It is frequently used
in the MD community in order to identify the linear subspace in which
the largest-amplitude motions occur, with the hope that these large-amplitude
motions are most informative of functionally relevant transitions
\cite{Amadei_Proteins17_412}.

Let $\mathbf{r}\in\mathbb{R}^{d}$ be a vector of order parameters
used, for example distances or Cartesian positions. Without restriction
of generality we assume that $\mathbf{r}$ is mean-free, i.e. the
mean of the data has already been subtracted. Note that $\mathbf{r}$
is generally only a subset of the full phase space coordinates, thus
$\mathbb{R}^{d}$ is a subset of $\Omega$. For example, in protein
simulation $\mathbf{r}$ usually only contains the protein coordinates,
but not those of the solvent. The covariance matrix $\mathbf{C}^{r}$
of the order parameters $\mathbf{r}$ is defined by the elements:
\[
c_{ij}^{r}=\langle r_{i}r_{j}\rangle
\]
while the estimator for trajectory data containing $N$ discrete time
steps is:
\[
\hat{c}_{ij}^{r}=\frac{1}{N-1}\sum_{t=1}^{N}r_{i}(t)r_{j}(t)
\]
The elements $c_{ij}^{r}$ are covariances between different order
parameters if $i\neq j$ and autocovariances if $i=j$. 

Principal components (PCs) are uncorrelated variables $\mathbf{y}$
that are obtained \emph{via} an orthonormal transform of the original
order parameters $\mathbf{r}$. For this, the eigenvectors $\mathbf{w}_{i}$
of the correlation matrix are obtained:
\[
\mathbf{C}^{r}\mathbf{w}_{i}=\mathbf{w}_{i}\sigma_{i}^{2},
\]
in matrix form, with the eigenvector matrix $\mathbf{W}=[\mathbf{w}_{1},...,\mathbf{w}_{d}]$
and the matrix of variances $\boldsymbol{\Sigma}^{2}=\mathrm{diag}(\sigma_{1}^{2},..,\sigma_{d}^{2})$.
\[
\mathbf{C}^{r}\mathbf{W}=\mathbf{W}\boldsymbol{\Sigma}^{2}.
\]
In order to transform an original coordinate vector $\mathbf{r}$
into principal components, we perform:
\begin{equation}
\mathbf{y}^{T}=\mathbf{r}^{T}\mathbf{W}\label{eq_pca-transform}
\end{equation}
Usually, principal components are sorted according to their autocovariance
$\sigma_{i}^{2}$. If $\sigma_{i}^{2}$ decays rapidly with $i$,
one often selects a threshold and ignores all PCs with smaller $\sigma_{i}^{2}$.
This is done by using some $\mathbf{W}'\in\mathbb{R}^{d\times m}$
matrix, which only contains the dominant $m<d$ column vectors of
$\mathbf{W}$. Used in this way, PCA is a tool for dimension reduction.
In the present paper we use PCA in two ways: (1) as a direct dimension
reduction tool to yield a subspace for clustering and subsequent Markov
model construction, and (2) to transform the original data into the
full set of principal components via Eq. (\ref{eq_pca-transform}),
thus arriving at a decorrelated coordinate set as an input for the
subsequent transform (see subsequent section). This way of using PCA
is called \emph{whitening} the data \cite{Karhunen::01}.

PCA is often used to analyze MD data, and has also been employed in
the construction of Markov models. We used PCA to reduce the dimension
of Pin WW protein simulations in order to build a protein folding
Markov model with a lagtime of only $\tau=2$ ns \cite{NoeSchuetteReichWeikl_PNAS09_TPT}.
Stock and co-workers have explored the possibility of using dihedral
angles as an input to PCA (dPCA). As angular coordinates, they cannot
be averaged in the same way like nonperiodic coordinates. Ref.\emph{
}\cite{Altis::08} thus suggests to use the sine and cosine of backbone
dihedral angles as input coordinates. It is found for hepta-alanin
that small-amplitude PCs are unimodal while large-amplitude PCs have
multimodal distributions and thus contain the interesting conformation
dynamics. Performing $k$-means clustering on the PCs did produce
states with high metastability. However, it has proven difficult to
analyze proteins with several secondary structure elements using dPCA.
In subsequent work \cite{Jain::10}, the approach was extended to
the ``dihedral PCA by parts'' method.

\subsection{Time-lagged independent component analysis (TICA)}

Like PCA, TICA \cite{Molgedey::94} uses a linear transform to map
the original order parameters $\mathbf{r}(t)$ to a new set of order
parameters $\mathbf{z}(t)$ --- the independent components (ICs).
Unlike PCs, ICs have to fulfill \emph{two} properties: 
\begin{enumerate}
\item they are uncorrelated and
\item their autocovariances at a fixed lag time $\tau$ are maximal. 
\end{enumerate}
The time-lagged covariance matrix $\mathbf{C}_{\tau}^{r}(\tau)$ is
defined by:
\[
c_{ij}^{r}(\tau)=\langle r_{i}(t)r_{j}(t+\tau)\rangle
\]
and the estimator for trajectory data containing $N$ time steps is
given by: 
\[
\hat{c}_{ij}^{r}(\tau)=\frac{1}{N-\tau-1}\sum_{t=1}^{N-\tau}r_{i}(t)r_{j}(t+\tau)
\]
The elements of $\mathbf{C}^{r}(\tau)$ are time-lagged autocovariances
if $i=j$ and time-lagged cross covariances if $i\neq j$. As shown
in the Appendix, this matrix is symmetric under the assumption of
reversible dynamics and in the limit of good statistics. For a finite
dataset, symmetricity must be enforced.

We seek a transformation matrix $\mathbf{U}=[\mathbf{u}_{1},...,\mathbf{u}_{d}]$
that diagonalizes $\mathbf{C}^{r}(0)$ (to fulfill property 1), and
maximizes the autocorrelations $c_{ii}^{z}(\tau)=\mathbf{u}_{i}^{T}\mathbf{C}^{r}(\tau)\mathbf{u}_{i}$
for every column $\mathbf{u}_{i}$ of $\mathbf{U}$ (to fulfill property
2). As described in Sec. \ref{sub_theory_optimal-linear-combination},
this is accomplished by solving:
\begin{equation}
\mathbf{C}^{r}(\tau)\mathbf{u}_{i}=\mathbf{C}^{r}(0)\mathbf{u}_{i}\hat{\lambda}_{i}(\tau),\label{eq_TICA-generalized-EVproblem}
\end{equation}
Eq. (\ref{eq_TICA-generalized-EVproblem}) is equivalent to Eq. (\ref{eq_Roothaan-Hall}).
See appendix for an illustrative derivation of (\ref{eq_TICA-generalized-EVproblem}).
As described in Sec. \ref{sub_theory_optimal-linear-combination},
the second-largest estimated eigenvalue is a lower bound for the real
second-largest propagator eigenvalue: $\hat{\lambda}_{2}(\tau)<\lambda_{2}(\tau)$.

ICs are now ordered according to the magnitude of the autocovariance
$\hat{\lambda}_{i}(\tau)$, and the IC's with the largest autocovariances
$\hat{\lambda}_{i}(\tau)$ will be called \emph{dominant}. Since the
dominant $m$ IC's yield the linear subspace in which most of the
slow processes are contained, it is reasonable to now perform a direct
clustering in this subspace, thus aiming at approximating the nonlinear
behavior of the slowest $m$ eigenfunctions with step functions. This
will yield a better approximation to the $m$ slowest timescales.
Rewriting Eq. (\ref{eq_TICA-generalized-EVproblem}) in matrix form,
with and the matrix of autocorrelations $\hat{\boldsymbol{\Lambda}}(\tau)=\mathrm{diag}(\hat{\lambda}_{1}(\tau),..,\hat{\lambda}_{d}(\tau))$
yields: 
\begin{equation}
\mathbf{C}^{r}(\tau)\mathbf{U}=\mathbf{C}^{r}(0)\mathbf{U}\hat{\boldsymbol{\Lambda}}(\tau).\label{eq_TICA-generalized-EVproblem-2}
\end{equation}
In order to transform an original coordinate vector $\mathbf{r}$
into independent components, we perform:
\begin{equation}
\mathbf{z}^{T}=\mathbf{r}^{T}\mathbf{U}\label{eq_tica-transform}
\end{equation}

How can (\ref{eq_TICA-generalized-EVproblem-2}) be solved? If $\mathbf{C}^{r}(0)$
or $\mathbf{C}^{r}(\tau)$ were invertible the generalized eigenvalue
problem could be transformed into a normal eigenvalue problem. But
as we expect some of our original order parameters to be highly correlated,
the determinants of $\mathbf{C}^{r}$ and $\mathbf{C}^{r}(\tau)$
will be nearly zero, prohibiting this option. Alternatively, one can
seek the solution of (\ref{eq_TICA-generalized-EVproblem}) via generalized
eigensolvers. 

However, there is a simple and efficient alternative to this: Problem
(\ref{eq_TICA-generalized-EVproblem-2}) can also be solved by solving
two simple eigenvalue problems using the AMUSE algorithm \cite{Tong::90}.
It consists of the following steps:
\begin{enumerate}
\item Use PCA to transform mean-free data $\mathbf{r}(t)$ into principal
components $\mathbf{y}(t)$. 
\item Normalize principal components: $\mathbf{y}'(t)=\boldsymbol{\Sigma}^{-1}\mathbf{y}(t)$. 
\item Compute the symmetrized time-lagged covariance matrix ${\mathbf{C}'}_{\tau}^{y'}=\frac{1}{2}[\mathbf{C}_{\tau}^{y'}+(\mathbf{C}_{\tau}^{y'})^{\dagger}]$
of the normalized PCs.
\item Compute an eigenvalue decomposition of ${\mathbf{C}'}_{\tau}^{y'}$,
obtaining eigenvector matrix $\mathbf{V}$ and project the trajectory
$\mathbf{y'}(t)$ onto the dominant eigenvectors to obtain $\mathbf{z}(t)$.
\end{enumerate}
This only works when the eigenvectors of ${\mathbf{C}'}_{\tau}^{y'}$
are uniquely defined, i.e. if the eigenvalues are not degenerated
\cite{Karhunen::01}. The main idea of this algorithm is, that properties
(1) and (2) can be fulfilled one after the other. First, steps 1 and
2 use PCA to produce decorrelated and normalized trajectories $\mathbf{y}'(t)$,
also known as whitening the data. Then steps 3 and 4 maximize the
time lagged autocovariances. Because the matrix $\mathbf{V}$ which
is used in step 4 is unitary, it preserves scalar products between
the vectors $\mathbf{y}'(t)$. Now if $\mathbf{y}'(t)$ are chosen
to be uncorrelated (and properly normalized) then also $\mathbf{z}(t)$
will be uncorrelated. 

In summary, the transformation Eq. (\ref{eq_tica-transform}) can
be written as a concatenation of three linear transforms:
\begin{equation}
\mathbf{z}^{T}(t)=\mathbf{r}^{T}(t)\mathbf{U}=\mathbf{r}^{T}(t)\mathbf{W}\boldsymbol{\Sigma}^{-1}\mathbf{V}.\label{eq_tica-transform-1}
\end{equation}
TICA will be used as a dimension reduction technique. Only the dominant
TICA components will be used to construct a Markov model.

\subsection{Markov model construction}

Markov models are constructed by first performing a data clustering
using an appropriate metric described in the results section (using
the EMMA command mm\_cluster) and subsequently converting the molecular
dynamics trajectory files into discrete trajectory files containing
the sequence of cluster indexes visited (using the EMMA command mm\_assign).
For the sake of the current paper, the main analysis is the behavior
of the relaxation timescales that are implied by the estimated Markov
model (EMMA command mm\_timescales).

All Markov model estimation is done as proposed in \cite{PrinzEtAl_JCP10_MSM1},
using the maximum probability estimator of reversible transition matrices
with a weak neighbor prior count matrix (EMMA default). Let us call
the transition matrix $\mathbf{T}(\tau)$, then it has the right eigenvectors
$\boldsymbol{\psi}_{i}$, the left eigenvectors $\boldsymbol{\phi}_{i}$
and the eigenvalues $\tilde{\lambda}_{i}$ according to the following
Eigenvalue equations:
\begin{eqnarray*}
\mathbf{T}(\tau)\boldsymbol{\psi}_{i} & = & \tilde{\lambda}_{i}(\tau)\boldsymbol{\psi}_{i}\\
\boldsymbol{\phi}_{i}^{T}\mathbf{T}(\tau) & = & \tilde{\lambda}_{i}(\tau)\boldsymbol{\phi}_{i}^{T}
\end{eqnarray*}
We order eigenvalues by descending norm. When $\mathbf{T}(\tau)$
is connected (irreducible), it will have a unique eigenvalue of norm
1. The corresponding eigenvector can be normalized to yield the stationary
distribution $\boldsymbol{\pi}$:
\[
\boldsymbol{\pi}^{T}=\boldsymbol{\pi}^{T}\mathbf{T}(\tau).
\]

Since $\mathbf{T}(\tau)$ fulfills detailed balance, the left and
right eigenvectors are related by:
\[
\boldsymbol{\phi}_{i}=\mathrm{diag}(\boldsymbol{\pi})\boldsymbol{\psi}_{i}
\]
The estimated (implied) relaxation timescales of the Markov model
are given by
\[
\tilde{t}_{i}=-\frac{\tau}{\mathrm{ln}\:\tilde{\lambda}(\tau)}.
\]
which are - ignoring statistical errors - related to the true relaxation
timescales by $\tilde{t}_{i}<t_{i}$ (see theory section), and are
typically larger than the timescales implied by the TICA eigenvalues
(see theory section and Fig. \ref{fig_approximation-scheme}).

\subsection{Optimal indicators}

Given the final Markov model transition matrix $\mathbf{T}(\tau)$,
we can now establish a simple way to quantify how well each of the
order parameters $r_{k}$ serving as an input serves as an indicator
of the slow process described by the eigenvector $\boldsymbol{\psi}_{i}$:
We simply compute the correlation between all pairs of order parameters
and eigenvectors, and then, for each eigenvector, choose those order
parameters that have a maximum correlation:
\begin{equation}
r_{\mathrm{opt}}(i)=\arg\max_{r_{k}}\frac{\langle r_{k}\boldsymbol{\psi}_{i}\rangle-\langle r_{k}\rangle\langle\boldsymbol{\psi}_{i}\rangle}{\sqrt{\langle r_{k}^{2}\rangle\langle\boldsymbol{\psi}_{i}^{2}\rangle}}.\label{eq_optimal-indicator}
\end{equation}
The averages in Eq. (\ref{eq_optimal-indicator}) can be computed
either \emph{via} Markov model states (in this case the average value
of $r_{k}$, is computed for every microstate $j$, obtaining $\bar{r}_{k,j}$,
and the correlation is given by $\langle r_{k}\boldsymbol{\psi}_{i}\rangle=\sum_{j=1}^{n}\pi_{j}\bar{r}_{k,j}\psi_{i,j}$).
Here we instead choose to evaluate the averages as a time average
over all trajectory data. In this case, the eigenvector coordinate
is given by the microstate each trajectory frame is associated to.

\section{Results}

The proposed methodology is demonstrated on two different peptide
systems: the fluorescent peptide MR121-GSGSW and the 30-residue natively
unstructured peptide KID.

MR121-GSGSW is a well-studied fluorescent peptide that has been extensively
characterized by experiments \cite{NeuweilerEtAl_JMB07}, simulations
\cite{Daidone_PlosOne10_MR121Kinetics} and also Markov models \cite{NoeDaidoneSmithAmadei_JPCB08_QGE,NoeEtAl_PNAS11_Fingerprints}.
Here, a data set of two explicit solvent simulations of 3 $\mu s$
each is used that is publicly accessible as a benchmark dataset for
the EMMA software package (see http://simtk.org/home/emma). The details
of the simulation setup are described in \cite{PrinzEtAl_JCP10_MSM1}.

The slowest relaxation timescale of the MR121-GSGSW data set has been
estimated to be between 20 and 30 ns, and it has been found that the
slowest processes are dominated by the interaction between MR121 and
the tryptophan residue \cite{NoeEtAl_PNAS11_Fingerprints}. The data
set is used as a benchmark system to test whether Markov model construction
in PCA or TICA coordinates manage to identify the slow parameters,
approximate the slow processes, and assign the correct timescales.

\begin{figure}
\includegraphics[width=1\columnwidth]{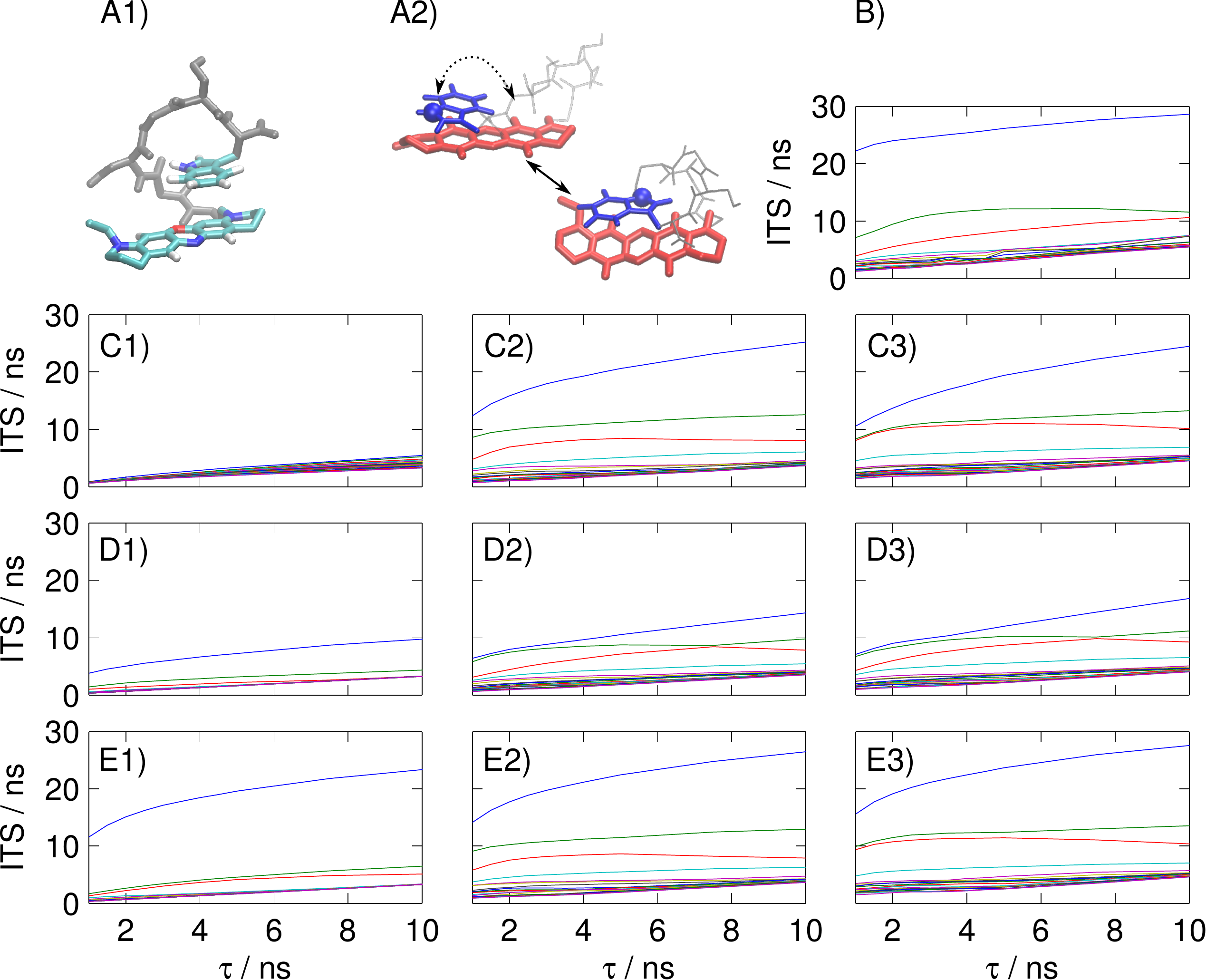}

\caption{\selectlanguage{english}%
\label{fig_1}\foreignlanguage{american}{MR121-GSGSW peptide and its
dominant relaxation timescales calculated \emph{via} different Markov
model construction methods. (A1) Sample structure of the peptide.
(A2) Illustration of the Trp coordinates used. The center position
of the Trp and the orientation vectors are given in a coordinate system
defined by the MR121 principal axes. (B) Relaxation timescales using
regular space RMSD clustering with approximately 1000 clusters. (C-E)
Relaxation timescales using $k$-means with 1000 clusters and Euclidean
metric but operating on different subspaces: (C1) Intramolecular distances
between all $C_{\alpha}$'s and ring centers. (C2) Center position
and orientation coordinates of the Trp moiety in the MR121 coordinate
system. (C3) Combined coordinate set including intramolecular distances
and Trp coordinates. (D1-3) Dominant PCA subspace of the combined
coordinate set using 1, 4, and 10 dimensions. (E1-3) Dominant TICA
subspace of the combined coordinate set using 1, 4, and 10 dimensions.}\selectlanguage{american}%
}
\end{figure}

\begin{figure}
\textbf{}%
\begin{minipage}[b][11cm][t]{0.5cm}%
\textbf{\Large A}{\Large \par}

\vspace{10cm}%
\end{minipage}\textbf{\includegraphics[width=0.9\columnwidth]{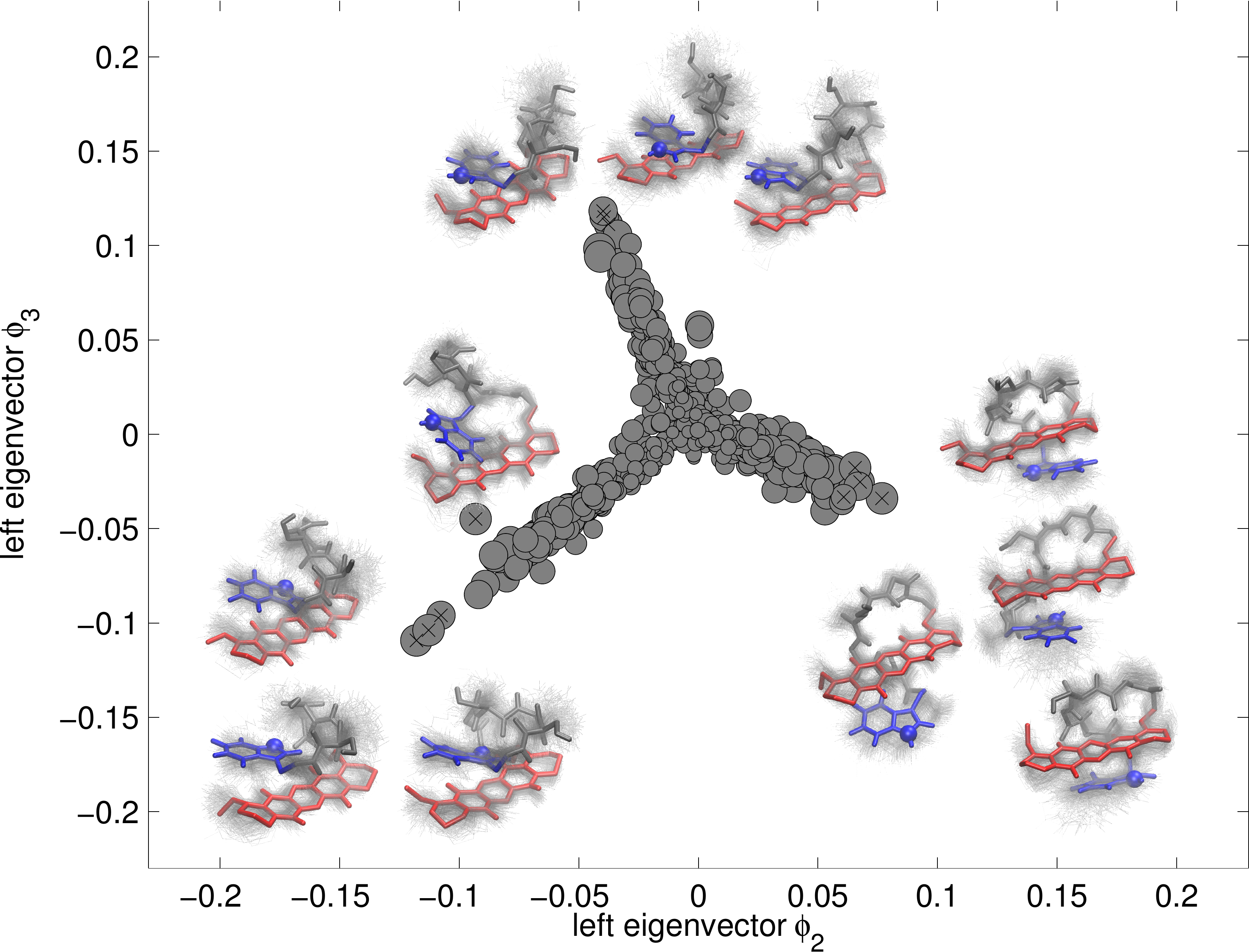}}

\textbf{\includegraphics[width=1\columnwidth]{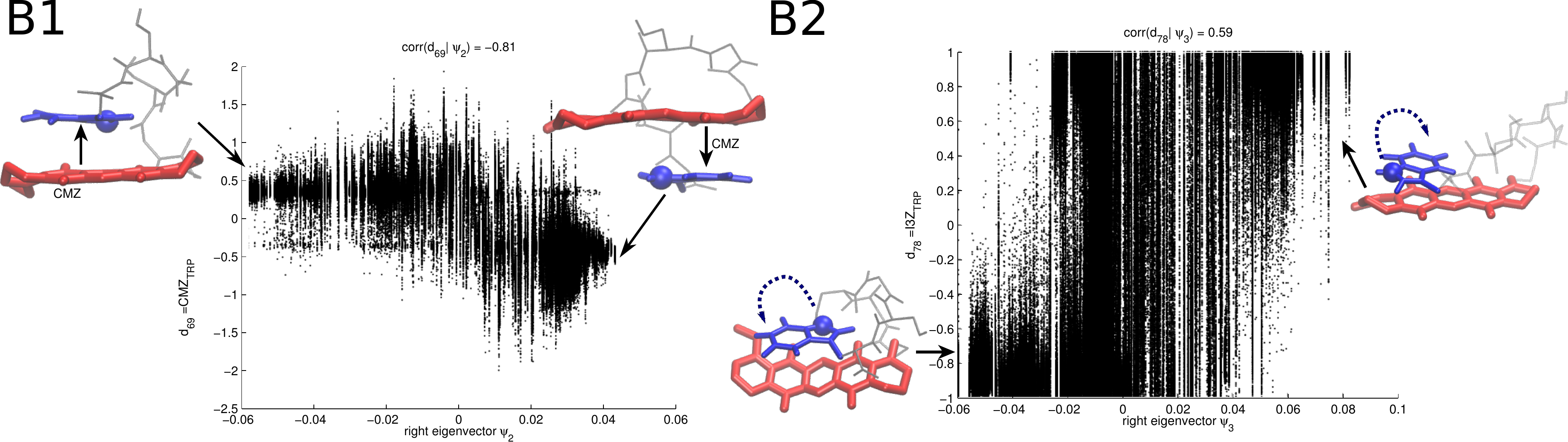}}

\caption{\selectlanguage{english}%
\label{fig_2}\foreignlanguage{american}{(A) \emph{Kinetic map} of
the two slowest relaxation processes of MR121-GSGSW (around 27 ns
and 13 ns) calculated from the Markov model shown in Fig \ref{fig_1}E3.
The grey discs mark the coordinates of the 1000 microstates in the
space of the left eigenvectors $\boldsymbol{\phi}_{2}$, $\boldsymbol{\phi}_{3}$.
The slowest relaxation of the system thus takes place on the horizontal
axis, the second-slowest one on the vertical axis, and distances are
associated with kinetic separation. The area of a disc is proportional
to the stationary probability of the corresponding microstate. Some
representative (kinetically distant and populous) microstates are
shown as molecular structures. (B) \emph{Optimal indicators} of the
slow processes. The scatter plots show the correlation between the
second and third right Markov model eigenvectors $\boldsymbol{\psi}_{2}$,
$\boldsymbol{\psi}_{3}$ and the order parameters most correlated
with them. The arrows in the structures show the optimal indicators.
(B1) The Trp $z$-position mediates the stacking order exchange and
has a correlation coefficient of 0.84 with the second eigenvector
$\boldsymbol{\psi}_{2}$ (timescale 27 ns). (B2) The smallest Trp
axis of inertia mediates the rotation of the side-chain and has a
correlation coefficient of 0.59 with the third eigenvector $\boldsymbol{\psi}_{3}$
(timescale 13 ns).}\selectlanguage{american}%
}
\end{figure}

Fig. \ref{fig_1}A1 shows a sample structure of MR121-GSGSW. Fig \ref{fig_1}B
shows a benchmark for the relaxation timescales computed by a regular-space
clustering in pairwise minimal RMSD metric using 1000 cluster centers.
The slowest processes are found at about 25 ns, 12 ns and 8 ns, slightly
larger---and thus more accurate according to the variational principle
in Eq. (\ref{eq_t2-approximation})---than by the coarser Markov model
in \cite{NoeEtAl_PNAS11_Fingerprints}. To set up the direct clustering,
two internal coordinate sets are considered: (i) the set of 66 distances
between 12 coordinates defined by the 5 $C_{\alpha}$'s and the 7
ring centers involved, and (ii) the center position and the orientation
vector coordinates of the tryptophan sidechain in a coordinate set
defined by the MR121 principal axes (see Fig. \ref{fig_1}A2 for an
illustration). Fig. \ref{fig_1}c1-3 show the results of direct k-means
clustering with 1000 cluster centers in the space of 66 intramolecular
distances (C1), only the 9 tryptophan coordinates (C2), and the combined
set (C3). It is clearly seen that the intramolecular distances are
not suited to resolve the slowest processes, while the tryptophan
coordinates resolve them very well. This can be understood from the
structural arrangements shown in Fig. \ref{fig_2} that are dominated
by the relative orientation of the tryptophan sidechain with respect
to the MR121 ring system. Especially the slowest process, the stacking-order
exchange of the two ring systems, cannot be well described by the
intramolecular distances that are similar when the tryptophan is ``above''
or ``below'' the MR121. Fig. \ref{fig_1}C3 shows that discretizing
the combined coordinate set resolves the slowest processes with similar
timescales as in the 9 Trp-coordinate set alone. This is not always
expected, as increasing the dimensionality of the space to be clustered
while keeping the number of clusters constant will often reduce the
resolution. 

In the subsequent PCA and TICA analysis different linear subspaces
of the combined coordinate set were considered. Interestingly, clustering
the principal components reduces the quality of the Markov model significantly.
This is explained by the fact that the largest-amplitude motions in
the present system is the transition between structures in which Trp
and MR121 are in contact, and open structures. However, open structures
have a very low population, giving rise to a rather fast timescale
of the opening/closing process. The slowest processes, involving different
arrangements and orientations of the Trp and MR121 while being in
contact, give rise to comparatively small amplitude motions. Using
one and four PCA components (D1 and D2), the three slowest processes
are not found. Using ten PCA components, the two slowest processes
are found, although slightly underestimated, while the third-slowest
process is not found.

Fig. \ref{fig_1}E1-3 shows that the TICA coordinates perform indeed
very well. Using only the single slowest TICA coordinate does resolve
the slowest process well and gives rise to a timescale of 20-25 ns,
close to the expected value. Using the four slowest TICA coordinates
resolves the two slowest processes well, while somewhat underestimating
the third process. With ten TICA coordinates all slow processes are
well resolved, and the timescales are found to be 27 ns, 13 ns, and
10 ns at a lagtime of $\tau=$10 ns---slightly larger than in any
of the other choices of metrics.

Fig. \ref{fig_2}A illustrates the structural transition involved
in the two slowest processes occurring at around 27 and 13 ns computed
from the ten-dimensional TICA Markov model. We display the 1000 microstates
in a visualization that we shall call \emph{kinetic map}, where the
coordinates are given by the two slowest left eigenvectors $\boldsymbol{\phi}_{2}$,
$\boldsymbol{\phi}_{3}$. For example, a cluster $i$ is drawn at
a position $(\phi_{2,i},\phi_{3,i})$ with a size proportional to
its stationary probability $\pi_{i}$. The map is termed ``kinetic'',
because similar positions in eigenvector spaces mean that the states
can relatively quickly reach one another, while distant positions
only exchange on timescales $t_{2}$ on the horizontal, and on timescale
$t_{3}$ on the vertical axis. The left eigenvectors are chosen instead
of the right eigenvectors, because the left eigenvectors are weighted
by the stationary distribution: $\phi_{k,i}=\pi_{i}\psi_{k,i}$. Thus,
points on the border of the map tend to have larger stationary probability.
Therefore, the extremal points are at the same time populous and kinetically
distant, and can roughly be associated with the most stable ``free
energy minima'', while the smaller clusters connecting them correspond
to transition states. The structures, shown for the most populous
and kinetically distinct clusters, indicate that the slowest relaxations
are associated with a stacking-order exchange of the MR121 and Trp
groups, and a rotation of the Trp group with respect to the MR121
group (see ``marker'' atom shown as a blue sphere).

Fig. \ref{fig_2}B illustrates the optimal indicators of the slowest
processes, i.e. the input order parameters that have the largest correlation
with the individual right Markov model eigenvectors $\boldsymbol{\psi}_{2}$
and $\boldsymbol{\psi}_{3}$. The correlation plots show that the
respective order parameters attain clearly different values at the
end-states of the transition, i.e. for the minimal and maximal values
of the respective eigenvector. At intermediate values of the eigenvectors,
i.e. transition states, the order parameter can access many different
values. This is easily seen in the slowest process (Fig. \ref{fig_2}b1),
where the best indicator is the Trp $z$-position that mediates the
stacking order exchange (correlation coefficient 0.84 with the second
eigenvector $\boldsymbol{\psi}_{2}$). While the value of the Trp
$z$-position is clearly defined in the transition end-states, where
the Trp is located ``above'' and ``below'' the MR121 moiety, the
transition states include open configurations where the Trp and the
MR121 are not in contact at all, and therefore all values of the $z$-position
are accessible in these states. A similar behavior is seen for the
second-slowest process (Trp sidechain rotation).

We now turn to another molecular system. Here, an extensive set of
simulations of the kinase inducible domain (KID) in explicit solvent
were investigated. KID is part of the cAMP response element-binding
protein (CREB). CREB is a transcription factor involved in processes
as important as glucose regulation and memory, and it binds the CREB-binding
protein (CBP), a well-known cancer-related molecular hub with around
300 interacting protein partners \cite{Kasper_2006}. KID belongs
to a large and important class of \emph{intrinsically unstructured
}peptides (IUP), encompassing many hormones, domains and even whole
proteins \cite{Uversky_2011}. Unstructured regions perform their
function even though they lack a well-defined secondary or tertiary
structure in solution. Although standardized algorithms exist to detect
unstructured regions on the basis of the primary amino acid sequence,
the structural details of how disordered regions exert their function
is still elusive. For example, some unstructured domains, including
KID, become folded upon binding \cite{Sugase_Dyson_Wright_2007};
it is therefore of much interest (e.g. for the druggability of protein-protein
interactions) to investigate whether the presence of pre-formed elements
causes folded conformations to be selected from the ensemble (conformational
selection) \cite{Mohan_Uversky_2006}, or whether the binding rather
occurs through induced-fit mechanics \cite{Shoemaker_Portman_Wolynes_2000}. 

To shed light on this problem, we set-up an ensemble of all-atom simulations
of the phosphorylated KID (pKID) domain. We have performed 7706 all-atom
explicit-solvent simulations of 24 ns each using the ACEMD software
\cite{Harvey_Giupponi_Fabritiis_2009}  on the GPUGRID distributed
computing platform \cite{Buch_Harvey_Giorgino_Anderson_DeFabritiis_2010},
yielding a total 168 $\mu s$ simulation data. However, due to the
short simulations, only short lagtimes could be used, presenting a
challenge to the Markov model construction. The detailed simulation
setup is described in the Appendix.

Fig. \ref{fig_3} shows the performance of different metrics in their
ability to resolve the slowest processes of KID. KID is a more difficult
case than the MR121-GSGSW peptide because its natively unstructured
nature gives rise to many fast large-amplitude motions which will
conceal the slow processes in most \emph{ad hoc} metrics. Fig. \ref{fig_3}B
and C show that neither regular-space clustering in minimal pairwise
RMSD metric, nor direct clustering in all distances yield a converged
estimated of the timescales up to lagtimes of 10 ns. Between these
two, regular-space RMSD is better, reaching a timescale of about 170
ns at $\tau=10\: ns$, while the direct clustering produces a timescale
below 100 ns at $\tau=10\: ns$. Higher choices of lagtimes were avoided
as they lead to a severe reduction of the usable data because the
connected set of clusters drops significantly below 100\% after that
point. Fig. \ref{fig_3}D1-3 show that the performance of principal
components is even worse than direct clustering, giving rise to timescale
estimates below 20 ns for one principal component and below 50 ns
for ten principal components. This confirms that the largest-amplitude
motions are not the slowest in KID.

Fig. \ref{fig_3}E1-3 show the performance of the TICA coordinate
using one, four, and ten dimensions. Using only the slowest TICA coordinate,
a slow process of >200 ns is found, that has not been resolved by
the clustering in any of the other metrics, however this timescale
does not converge for lagtimes up to 10 ns. Using only the four slowest
TICA coordinates, there are already three processes resolved that
are above 100 ns, and the convergence behavior improves. Using the
ten slowest TICA coordinates, five processes slower than 100 ns are
resolved. The slowest process converges to a timescale around 220
ns and does so already at a lagtime $\tau$ of 2-5 ns. Thus, the lagtime
needed is a factor of 50-100 smaller than the timescale of the process,
indicating a very good discretization of the corresponding process.

Fig. \ref{fig_4}A illustrates the structural transitions associated
with the two slowest relaxation processes of KID as identified by
the Markov model using ten-dimensional TICA model. We have decided
to focus on the two slowest processes around 200 and 220 ns relaxation
time, because they are somewhat separated from the next-slowest processes
occurring at around 100 ns. As the peptide has great structural variability
it is of little value to plot all relevant structures. Therefore,
we have plotted the positions of the microstates again in a kinetic
map, using the coordinates of the two dominant left eigenvectors $\boldsymbol{\phi}_{2}$,
$\boldsymbol{\phi}_{3}$. It is seen that at the slowest timescales,
the system rearranges between mostly open and disordered structures
(left), structures with one helix folded or partially folded (top
right), and hairpin-like structures (bottom right). Thus the system
has some residual helical structure, although it is not very stable
in absence of a stabilizing binding partner. 

Fig. \ref{fig_4}B illustrates the optimal indicators of the slowest
processes, i.e. the input order parameters that have the largest correlation
with the resulting right Markov model eigenvectors $\boldsymbol{\psi}_{2}$
and $\boldsymbol{\psi}_{3}$. Like for MR121-GSGSW, the correlation
plots show that the respective order parameters are mainly able to
distinguish the end-states of the transition, but unlike for MR121-GSGSW,
multiple $C_{\alpha}-C_{\alpha}$ distances are almost equally good
indicators for the same process. Fig. \ref{fig_4}B shows correlation
plots of the best indicators of $\boldsymbol{\psi}_{2}$ and $\boldsymbol{\psi}_{3}$,
but indicates the five best correlations (all correlation coefficients
above 0.7) in the structure. It is seen that the slowest process (timescale
220 ns) is best described by a hinge opening and closing, where the
closed hinge appears to induce at least partial formation of the N-terminal
helix (red, see Fig. \ref{fig_4}B2). This is consistent with NMR
experiments that have shown the N-terminal region to be approximately
50\% helical in the apo-form \cite{RadhakrishnanWrite_Cell97_KIDKIX}.
The second-slowest process (timescale 200 ns) is best described by
partial helix formation of the C-terminal part (blue) of the protein.

\section{Discussion}

In the present manuscript we have derived a method to find the optimal
linear combination of input coordinates for approximating the slowest
relaxation processes in complex conformational rearrangements of molecules.
It is shown that an implementation for this method is already known
in statistics as the TICA method, which is combined here with Markov
modeling in order to construct models of the slow relaxation processes
and precise estimates of the related relaxation timescales. It is
shown that this approach of constructing Markov models yields slower
timescales, and thus a more precise approximation to the true relaxation
processes, than previous approaches. This is also achieved for the
natively unstructured peptide KID where established approaches such
as direct clustering in distance space, minimal-RMSD-based clustering,
or clustering in PCA space did not perform well because the largest-amplitude
motions were not good indicators of the slowest relaxation processes.

Beyond having an approach to construct quantitatively accurate Markov
models in a way that is more robust than most previous approaches,
we readily obtain a way to find \emph{best indicators} of the slowest
transitions. Best indicators are those molecular order parameters
that are best correlated with the Markov model Eigenvectors describing
the slowest processes, and thus serve as candidates for good reaction
coordinates. Being able to point out such indicators provides a way
to make the sometimes complex structural rearrangements readily understandable.

\section*{Acknowledgements}

We thank all the volunteers of GPUGRID who donated GPU computing time
to the project. We are grateful to Thomas Weikl (MPI Potsdam) for
advice and support. G.P.-H. acknowledges support from German Science
Foundation DFG fund NO 825-3. F.P. acknowledges funding from the Max
Planck society. T.G. gratefully acknowledges former support from the
\textquotedblleft{}Beatriu de Pinós\textquotedblright{} scheme of
the Agència de Gestió d\textquoteright{}Ajuts Universitaris i de Recerca
(Generalitat de Catalunya). G.D.F. acknowledges support from the Ramón
y Cajal scheme and support by the Spanish Ministry of Science and
Innovation (Ref. BIO2011-27450). F.N. acknowledges funding from DFG
center \textsc{Matheon}.

\clearpage

\clearpage

\clearpage

\begin{figure}
\includegraphics[width=1\columnwidth]{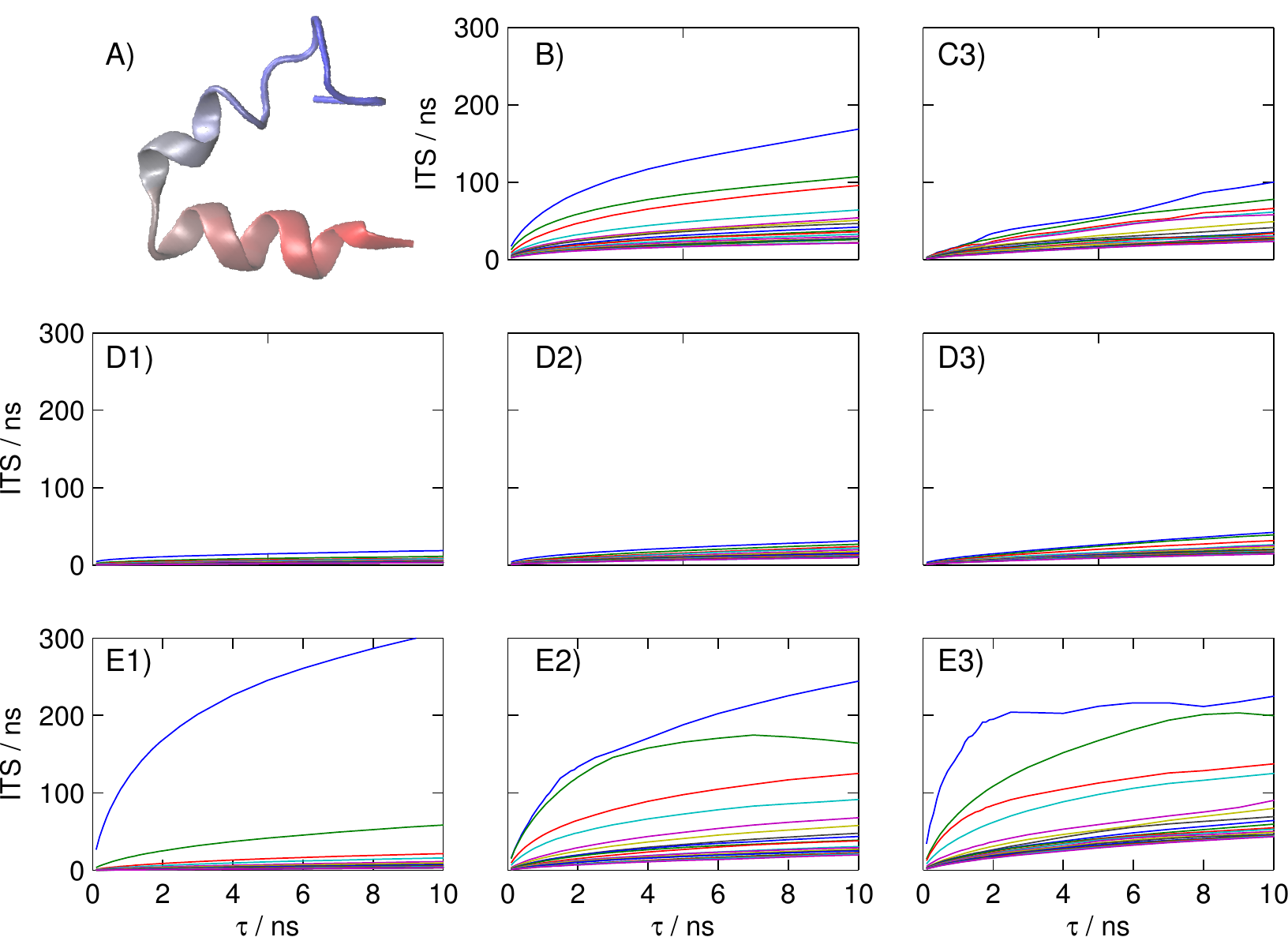}

\caption{\selectlanguage{english}%
\label{fig_3}\foreignlanguage{american}{KID peptide and its estimated
dominant relaxation timescales using different Markov model construction
methods. (A) Sample structure of KID. (B) Relaxation timescales using
regular space RMSD clustering with 1000 clusters. (C-E) Relaxation
timescales using $k$-means with 1000 clusters and Euclidean metric
but operating on different subspaces. (C) All $C_{\alpha}-C_{\alpha}$
distances. (D1-3) Dominant PCA subspace of $C_{\alpha}-C_{\alpha}$
distances using 1, 4, and 10 dimensions. (E1-3) Dominant TICA subspace
of $C_{\alpha}-C_{\alpha}$ distances using 1, 4, and 10 dimensions.}\selectlanguage{american}%
}
\end{figure}

\clearpage

\begin{figure}
\textbf{}%
\begin{minipage}[b][11cm][t]{0.5cm}%
\textbf{\Large A}{\Large \par}

\vspace{10cm}%
\end{minipage}\includegraphics[width=0.9\columnwidth]{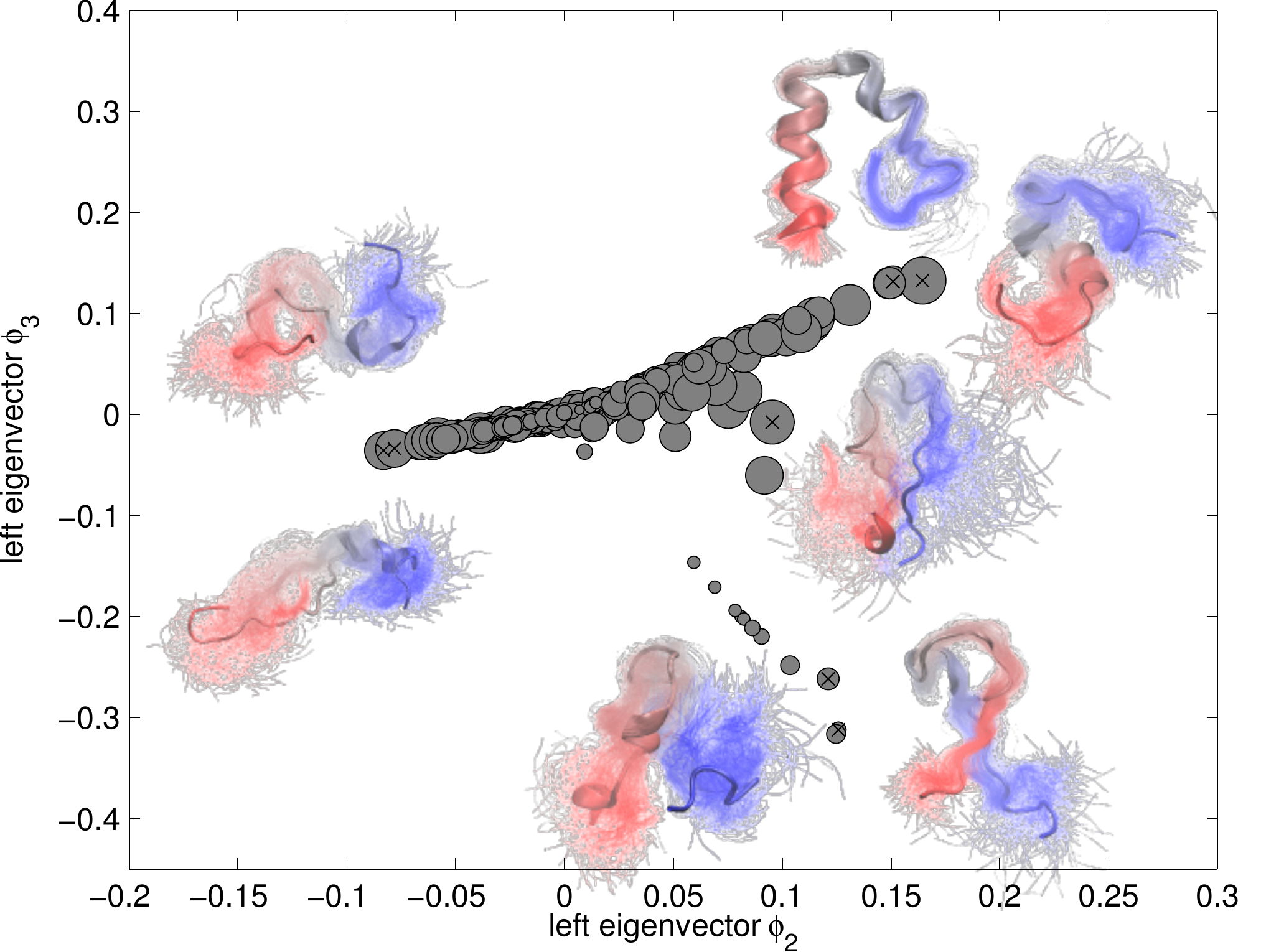}

\textbf{\includegraphics[width=1\columnwidth]{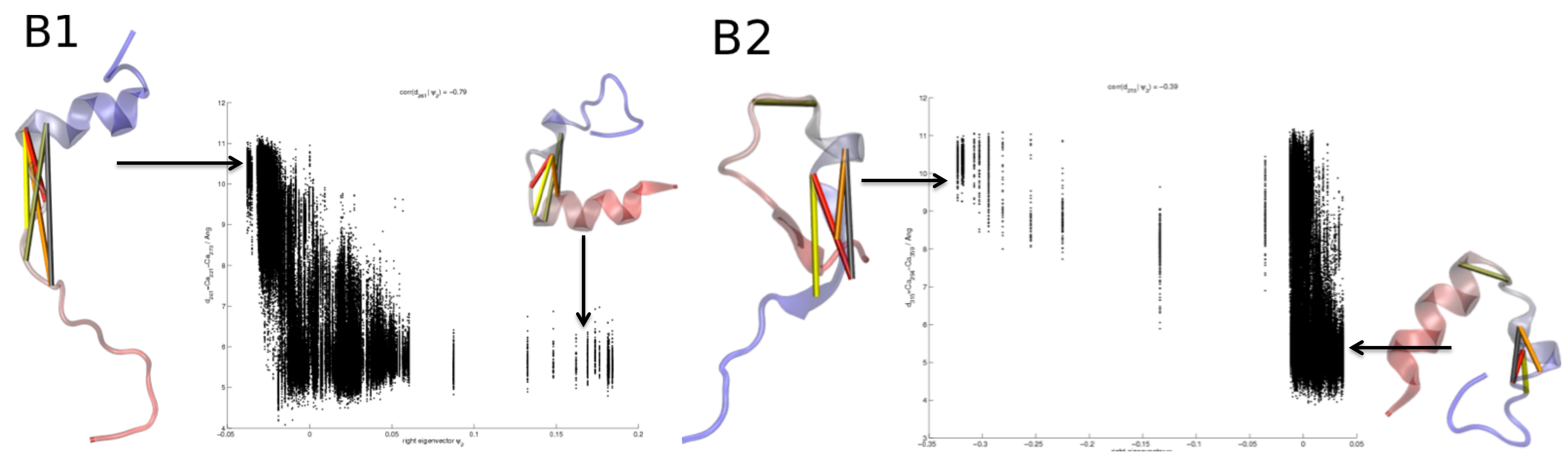}}

\caption{\selectlanguage{english}%
\label{fig_4}\foreignlanguage{american}{(A) \emph{Kinetic map} of
the two slowest relaxation processes of the KID peptide (around 200
ns and 220 ns) calculated from the Markov model shown in Fig \ref{fig_3}E3.
The grey discs mark the coordinates of the 1000 microstates in the
space of the left eigenvectors $\boldsymbol{\phi}_{2}$, $\boldsymbol{\phi}_{3}$.
The slowest relaxation of the system thus takes place on the horizontal
axis, the second-slowest one on the vertical axis, and distances are
associated with kinetic separation. The area of a disc is proportional
to the stationary probability of the corresponding microstate. Some
representative (kinetically distant and populous) microstates are
shown as molecular structures. (B) \emph{Optimal indicators} of the
slow processes. The scatter plots show the correlation between the
second and third right Markov model eigenvectors $\boldsymbol{\psi}_{2}$,
$\boldsymbol{\psi}_{3}$ and the order parameters most correlated
with them. The colored lines show all five best indicators, all having
correlation coefficients with the respective eigenvectors of 0.7 or
greater. The slowest process may thus be described as opening / closing
of the hinge between the two helical domains of KID (timescale 220
ns), while hinge-closing is associated with at least partial N-terminal
helix formation (red). The second-slowest process may be described
as partial helix formation in the ``blue'' region (timescale 200
ns).}\selectlanguage{american}%
}
\end{figure}

\clearpage

\section*{Appendix}

\subsubsection*{Derivation of TICA}

The generalized Eigenvalue problem of Eq. (\ref{eq_Roothaan-Hall}),
and more specifically the TICA problem can be derived in different
ways. It goes back to the classical Ritz method \cite{Ritz_JReineAngewMathe09_Variationsprobleme}
and can be found in many mathematical texts. In the following we sketch
a standard derivation using variational calculus, see also \cite{Jolliffe_PCA-book,HyvaerinenKarhunenOja_ICA_Book}
for a thorough discussion of this approach.

Let $\mathbf{r}\in\mathbb{R}^{d}$ be a vector of coordinates used,
for example distances or Cartesian positions. Without restriction
of generality we assume that $\mathbf{r}$ is mean-free, i.e. the
mean of the data has already been subtracted. Note that $\mathbf{r}$
is contains generally only a subset of the full phase space coordinates,
thus $\mathbb{R}^{d}$ is a subset of $\Omega$.

We now seek new coordinates $\mathbf{z}\in\mathbb{R}^{m}$ as a linear
transformation of $\mathbf{r}$ such that
\begin{enumerate}
\item $\mathbf{z}$ are uncorrelated
\item the autocovariances of $\mathbf{z}$ at a fixed lag time $\tau$ are
maximal. 
\end{enumerate}
We will show that if coordinates $\mathbf{z}$ are given by a weighted
sum of $\mathbf{r}$ 
\begin{equation}
z_{i}(\mathbf{r})=\sum_{k=1}^{d}u_{ik}r_{k}\label{eq_linear-combination-1}
\end{equation}
the weight coefficients have to fulfill the generalized eigenvalue
problem (see theory section)

\begin{equation}
\mathbf{C}^{r}(\tau)\mathbf{u}_{i}=\hat{\lambda}_{i}\mathbf{C}^{r}(0)\mathbf{u}_{i}\label{eq_Roothaan-Hall-1}
\end{equation}
where $\mathbf{C}_{\tau}^{r}(\tau)$ is the time-lagged covariance
matrix that is defined by:
\begin{equation}
c_{ij}^{r}(\tau)=\langle r_{i}(t)r_{j}(t+\tau)\rangle\label{eq_time-covariance}
\end{equation}
To prove this we rewrite the covariance matrix of $\mathbf{z}$ and
the time-lagged covariance matrix of $\mathbf{z}$ using the defining
equations (\ref{eq_linear-combination-1}), and (\ref{eq_time-covariance}).

\begin{eqnarray*}
c_{ij}^{z}(0) & = & \langle z_{i}(t)z_{j}(t)\rangle=\sum_{k,l}u_{ik}u_{jl}\langle r_{k}(t)r_{l}(t)\rangle=\sum_{k,l}u_{ik}u_{jl}c_{kl}^{r}(0)\\
c_{ij}^{z}(\tau) & = & \langle z_{i}(t)z_{j}(t+\tau)\rangle=\sum_{k,l}u_{ik}u_{jl}\langle r_{k}(t)r_{l}(t+\tau)\rangle=\sum_{k,l}u_{ik}u_{jl}c{}_{kl}^{r}(\tau)
\end{eqnarray*}

We wish to maximize $c_{ij}^{z}(\tau)$ (property 3) under the constraint,
that $c_{ij}^{z}(0)=\delta_{ij}$ (property 2). We start by computing
one coordinate $z_{1}$ with maximal autocovariance. It is given by
the weighted sum $z_{1}=\sum_{i,j}u_{i}u_{j}c_{ij}^{r}(0)$, where
we used the shorthand notation $u_{i}=u_{1i}$. The constraint (2)
for $z_{1}$ is now $c_{11}^{z}(0)=\sum_{i,j}u_{i}u_{j}c_{ij}^{r}(0)=1$.

Since the matrix-elements $c_{ij}^{r}(0)$ are fixed, the autocovariance
$c_{11}^{z}(\tau)$ can be treated as a differentiable function of
the coefficients $u_{i}$. Therefore, we need to maximize the function

\[
F(u_{1},\ldots,u_{d})=\left(\sum_{k,l}u_{k}u_{l}c{}_{kl}^{r}(\tau)\right)-\hat{\lambda}_{1}\left(\sum_{k,l}u_{k}u_{l}c_{kl}^{r}(0)-1\right)
\]
where $\hat{\lambda}$ is the Lagrange multiplier. We perform the
maximization by setting the partial derivatives of $F$ with respect
to the weight coefficients to zero.

\[
0=\frac{\partial F}{\partial u_{k}}=\left(\sum_{l}u_{l}c_{kl}^{r}(\tau)\right)-\hat{\lambda}_{1}\left(\sum_{l}u_{l}c_{kl}^{r}(0)\right)
\]
Rearranging and rewriting this equation in matrix-vector form leads
to (\ref{eq_Roothaan-Hall-1}) for $i=1$. The same argument is used
for the subsequent eigenvalues. We now prove that the solutions of
(\ref{eq_Roothaan-Hall-1}) fulfill the properties requested above:
\begin{enumerate}
\item \textbf{The IC's obtained by solving }(\ref{eq_Roothaan-Hall-1})\textbf{
are uncorrelated}: Let $\mathbf{u}_{i}$ be a generalized eigenvector
with eigenvalue $\lambda_{i}$ and let $\mathbf{u}_{j}$ be a generalized
eigenvector with eigenvalue $\lambda_{j}\neq\lambda_{i}$. Then the
orthogonality condition
\begin{equation}
\mathbf{u}_{i}^{T}\mathbf{C}^{r}(0)\mathbf{u}_{j}=\delta_{ij}\label{eq_orthogonality-condition}
\end{equation}
will hold if $\mathbf{C}^{r}(0)$ and $\mathbf{C}^{r}(\tau)$ are
symmetric matrices. If $\mathbf{u}_{i}$ and $\mathbf{u}_{j}$ are
used as the weights in (\ref{eq_linear-combination-1}) this is equivalent
to $c_{ij}^{z}(0)=\delta_{ij}$.\\
Proof:
\[
\lambda_{i}\mathbf{C}^{r}(0)\mathbf{u}_{i}\cdot\mathbf{u}_{j}=\mathbf{C}^{r}(\tau)\mathbf{u}_{i}\cdot\mathbf{u}_{j}=\mathbf{u}_{i}\cdot\mathbf{C}^{r}(\tau)\mathbf{u}_{j}=\mathbf{u}_{i}\cdot\lambda_{j}\mathbf{C}^{r}(0)\mathbf{u}_{j}=\lambda_{j}\mathbf{C}^{r}(0)\mathbf{u}_{i}\cdot\mathbf{u}_{j}
\]
Therefore $0=(\lambda_{i}-\lambda_{j})(\mathbf{u}_{i}^{T}\mathbf{C}^{r}(0)\mathbf{u}_{j})$.
Because $\lambda_{i}\neq\lambda_{j}$ the orthogonality condition
must hold. This does not hold, if eigenvectors are degenerate i.e.
$\lambda_{i}=\lambda_{j}$ for some $i$, $j$. However degeneracy
can be avoided by changing the lag time $\tau$ such that no eigenvalues
with large magnitude coincide. Solutions with smaller eigenvalues
might still be degenerate, but this is unproblematic since these solutions
are discarded for clustering. In addition, ``fast'' modes will necessarily
be uncorrelated with ``slow'' modes, because their eigenvalues are
far apart. 
\item \textbf{The autocovariances at a fixed lag time $\tau$ are maximal}:
\\
We show that the autocovariances are identical to the Lagrange multipliers,
and thus to the eigenvalues in Eq. \ref{eq_Roothaan-Hall-1}: 
\begin{equation}
c_{ij}^{z}(\tau)=\hat{\lambda}_{i}\delta_{ij}\label{eq_eigenvalue-relation}
\end{equation}
To see this, multiply (\ref{eq_Roothaan-Hall-1}) with $\mathbf{u}_{i}^{T}$
from the left, to obtain 
\[
c_{ji}^{z}(\tau)=\mathbf{u}_{j}^{T}\mathbf{C}^{r}(\tau)\mathbf{u}_{i}=\hat{\lambda}_{i}\mathbf{u}_{j}^{T}\mathbf{C}^{r}(0)\mathbf{u}_{i}
\]
now use the orthogonality condition (\ref{eq_orthogonality-condition})
\[
c_{ji}^{z}(\tau)=c_{ij}^{z}(\tau)=\mathbf{u}_{j}^{T}\mathbf{C}^{r}(\tau)\mathbf{u}_{i}=\hat{\lambda}_{i}\delta_{ij}
\]
To show that the optimum found is indeed a maximum, we calculate the
Hessian of the constrained autocovariance $c_{11}^{z}(\tau)$. Its
elements are:
\[
H_{kl}=\frac{\partial^{2}F}{\partial u_{k}\partial u_{l}}=c_{kl}^{r}(\tau)-\hat{\lambda}_{1}c_{kl}^{r}(0)
\]
and show that it is a positive definite matrix
\[
\mathbf{x}^{T}\mathbf{H}\mathbf{x}=\mathbf{x}^{T}\mathbf{C}^{r}(\tau)\mathbf{x}-\hat{\lambda}_{1}\mathbf{x}^{T}\mathbf{C}^{r}(0)\mathbf{x}<0\;\forall\mathbf{x}
\]
We first expand $\mathbf{x}$ in the basis of the generalized eigenvectors
$\mathbf{x}=\sum_{i}^{m}\mathbf{u}_{i}(\mathbf{u}_{i}\cdot\mathbf{x})=\sum_{i}^{m}\mathbf{u}_{i}c_{i}$
and use equations (\ref{eq_orthogonality-condition}) and (\ref{eq_eigenvalue-relation})
\[
\sum_{i,j}c_{i}c_{j}\mathbf{u}_{i}^{T}\mathbf{C}^{r}(\tau)\mathbf{u}_{j}-\hat{\lambda}_{1}\sum_{i,j}c_{i}c_{j}\mathbf{u}_{i}^{T}\mathbf{C}^{r}(0)\mathbf{u}_{j}=\sum_{i}c_{i}^{2}\hat{\lambda}_{i}-\hat{\lambda}_{1}\sum_{i}c_{i}^{2}
\]
Without loss of generality, we assume that the solution vectors of
Eq. (\ref{eq_Roothaan-Hall-1}) were sorted by descending eigenvalues
$\hat{\lambda}_{i}$ to obtain an ordering from ``slow'' modes to
``fast'' modes, $\hat{\lambda}_{1}>\hat{\lambda}_{2}>\ldots>\hat{\lambda}_{m}$.
From this follows $\sum_{i}c_{i}^{2}\hat{\lambda}_{i}-\hat{\lambda}_{1}\sum_{i}c_{i}^{2}\leq0$
for the first solution $\mathbf{u}_{1}$. The second solution is restricted
to a subspace that is orthogonal to $\mathbf{u}_{1}$ according to
(\ref{eq_orthogonality-condition}) and (\ref{eq_eigenvalue-relation})
\[
\mathbf{x}^{T}\mathbf{C}^{r}(0)\mathbf{u}_{1}=\mathbf{x}^{T}\mathbf{C}^{r}(\tau)\mathbf{u}_{1}=0
\]
Therefore we can ignore the coefficient $c_{1}$ in the development
of $\mathbf{x}$ and obtain the quadratic form 
\[
\sum_{i=2}c_{i}^{2}\hat{\lambda}_{i}-\hat{\lambda}_{2}\sum_{i=2}c_{i}^{2}
\]
Again, this is negative, because $\hat{\lambda}_{2}$ is the largest
eigenvalue in the sum. This procedure can be extended to the third,
fourth,... eigenvalue, showing that the optima are minima for all
solutions. 
\end{enumerate}
As a result, we can sort the solution vectors of Eq. (\ref{eq_Roothaan-Hall-1})
by descending eigenvalues $\hat{\lambda}_{i}$ to obtain an ordering
from ``slow'' modes to ``fast'' modes.

\subsubsection*{Symmetricity and Symmetrization of the time-lagged covariance matrix}

Consider the correlation matrix of mean-free coordinates $\mathbf{r}$
for lag time $\tau$:
\[
\mathrm{c}_{ij}^{r}(\tau)=\langle r_{i}(t)\: r_{j}(t+\tau)\rangle
\]
and the correlation matrix for lag time $\tau$:
\begin{eqnarray*}
\mathrm{cor}_{ij}^{r}(\tau) & = & \frac{\mathrm{c}_{ij}^{r}(\tau)}{\sigma_{i}\sigma_{j}}=\frac{\langle r_{i}(t)r_{j}(t+\tau)\rangle}{\sqrt{\langle r_{i}^{2}(t)\rangle\langle r_{j}^{2}(t)\rangle}}\\
 & = & \int_{x}\int_{y}dxdy\: xy\: p(r_{i}(t)=x,r_{j}(t+\tau)=y)
\end{eqnarray*}

where $p(x(t)=x,y(t+\tau)=y)$ is the unconditional transition probability
between the set $S_{1}=\{r_{i}=x\}$ and the set $S_{2}=\{r_{j}=y\}$
within time lag $\tau$. In statistically reversible dynamics, such
an unconditional set-transition probability is symmetric (this follows
directly from integrating the detailed balance condition $\mu(\mathbf{x})p_{\tau}(\mathbf{x}\mid\mathbf{y})=\mu(\mathbf{y})p_{\tau}(\mathbf{y}\mid\mathbf{x})$
over the sets). Thus, we can exchange time indexes and show:
\begin{eqnarray*}
\mathrm{cor}_{ij}^{r}(\tau) & = & \int_{x}\int_{y}dxdy\: xy\: p(r_{i}(t+\tau)=x,r_{j}(t)=y)\\
 & = & \int_{y}\int_{x}dydx\: yx\: p(r_{j}(t)=y,r_{i}(t+\tau)=x)\\
 & = & \mathrm{cor}_{ji}^{r}(\tau).
\end{eqnarray*}

And then trivially
\[
\mathrm{c}_{ij}^{r}(\tau)=\mathrm{c}_{ij}^{r}(\tau)\:\:\:\:\forall\tau
\]

When estimating correlation or covariance matrices from simulations,
one cannot expect $c_{ij}=c_{ji}$ to hold. A trivial method is to
use
\[
c_{ij}(\tau)=\frac{1}{2}(\hat{c}_{ij}(\tau)+\hat{c}_{ji}(\tau))
\]
where $\hat{c}_{ij}(\tau)$ is the simulation estimate.

\subsubsection*{Simulation setup, KID}

The coordinates of the phosphorylated KID domain (28 residues, CREB
residues 119-146) were extracted from chain B of the entry 1KDX deposited
in the Protein Data Bank. The entry represent the folded configuration
of the pKID-CBP bound structure, determined through NMR \cite{Sugase_Dyson_Wright_2007}.
Neutral acetylated and N-methyl caps were added to avoid artifactual
charges at the peptide's termini; the protein was solvated with 6572
water molecules and a 85 mM KCl concentration (matching the experimental
ionic strength). The system was then parametrized with the AMBER ff99SB-ILDN
forcefield \cite{Lindorff_2010}; water and ions were modeled respectively
with the TIP3P and Joung-Cheatham parameter sets \cite{Joung_Cheatham_2008,jorgensen_comparison_1983}.
The system thus prepared was first equilibrated for 24 ns in the constant-pressure
ensemble, during which it stabilized at a volume of approximately
60 Å$^{3}$. The peptide was then denatured by heating it at 500 ºK
for 17.6 ns in constant-volume conditions; 176 frames were extracted
from this trajectory and used as starting configurations for the production
runs. All of the simulations were performed with a time step of 4
fs, enabled by the hydrogen mass repartitioning scheme \cite{feenstra_improving_1999};
long-range electrostatic forces were computed with the particle-mesh
Ewald summation method. A nonbonded cutoff distance of 9 Å was used
with a switching distance of 7.5 Å for Van der Waals interactions,
while the lengths of bonds involving hydrogen atoms were constrained
with the SHAKE algorithm. 

A set of 7706 production runs was executed on the GPUGRID distributed
computing network \cite{Buch_Harvey_Giorgino_Anderson_DeFabritiis_2010}.
Each simulation was performed in the constant-volume ensemble at 315
ºK for 24 ns with the same parametrization used for equilibration,
storing structural snapshots every 100 ps. Each production simulation
begun either from one of the configurations visited during the denaturation
run, or frames visited during preceding production trajectories. Starting
frames were selected iteratively with an adaptive strategy in order
to minimize the statistical uncertainty on the largest eigenvalue,
computed on the already available simulation data, based on Singhal
and Pande's algorithm \cite{Hinrichs_Pande_2007}. 

\clearpage

\bibliographystyle{plain}

\begin{thebibliography}{10}

\bibitem{Karhunen::01}
{Aapo Hyv\"arinen}, {Juha Karhunen}, and {Erkki Oja}.
\newblock {\em Independent {C}omponent {A}nalysis}, chapter~18, page 344.
\newblock John Wiley \& Sons, 2001.

\bibitem{HyvaerinenKarhunenOja_ICA_Book}
Erkki~Oja Aapo~Hyv{\"a}rinen, Juha~Karhunen.
\newblock {\em Independent Component Analysis}.
\newblock John Wiley \& Sons, 2001.

\bibitem{Altis::08}
Alexandros Altis, Moritz Otten, Phuong~H. Nguyen, Rainer Hegger, and Gerhard
  Stock.
\newblock Construction of the free energy landscape of biomolecules via
  dihedral angle principal component analysis.
\newblock {\em The Journal of Chemical Physics}, 128(24):245102, 2008.

\bibitem{Amadei_Proteins17_412}
A.~Amadei, A.~B. Linssen, and H.~J.~C. Berendsen.
\newblock {Essential dynamics of proteins}.
\newblock {\em Proteins}, 17:412--225, 1993.

\bibitem{BeauchampEtAl_MSMbuilder2}
Kyle~A. Beauchamp, Gregory~R. Bowman, Thomas~J. Lane, Lutz Maibaum, Imran~S.
  Haque, and Vijay~S. Pande.
\newblock {MSMBuilder2: Modeling Conformational Dynamics at the Picosecond to
  Millisecond Scale.}
\newblock {\em Journal of chemical theory and computation}, 7(10):3412--3419,
  October 2011.

\bibitem{Beauchamp::12}
Kyle~A. Beauchamp, Robert McGibbon, Yu-Shan Lin, and Vijay~S. Pande.
\newblock Simple few-state models reveal hidden complexity in protein folding.
\newblock {\em Proceedings of the National Academy of Sciences}, 2012.

\bibitem{BerezhkovskiiHummerSzabo_JCP09_Flux}
Alexander Berezhkovskii, Gerhard Hummer, and Attila Szabo.
\newblock {Reactive flux and folding pathways in network models of
  coarse-grained protein dynamics.}
\newblock {\em The Journal of chemical physics}, 130(20), May 2009.

\bibitem{Biarnes_Pietrucci_Marinelli_Laio_2012}
Xevi Biarn\'es, Fabio Pietrucci, Fabrizio Marinelli, and Alessandro Laio.
\newblock Metagui. a vmd interface for analyzing metadynamics and molecular
  dynamics simulations.
\newblock {\em Computer Physics Communications}, 183(1):203--211, Jan 2012.

\bibitem{Bowman_JCP09_Villin}
Gregory~R. Bowman, Kyle~A. Beauchamp, George Boxer, and Vijay~S. Pande.
\newblock {Progress and challenges in the automated construction of Markov
  state models for full protein systems.}
\newblock {\em J. Chem. Phys.}, 131(12):124101+, September 2009.

\bibitem{Bowman::12}
Gregory~R. Bowman and Phillip~L. Geissler.
\newblock Equilibrium fluctuations of a single folded protein reveal a
  multitude of potential cryptic allosteric sites.
\newblock {\em Proceedings of the National Academy of Sciences},
  109(29):11681--11686, 2012.

\bibitem{BowmanVoelzPande_JACS11_FiveHelixBundle-TripletQuenching}
Gregory~R. Bowman, Vincent~A. Voelz, and Vijay~S. Pande.
\newblock {Atomistic Folding Simulations of the Five-Helix Bundle Protein
  Lambda 6-85}.
\newblock {\em Journal of the American Chemical Society}, 133(4):664--667,
  February 2011.

\bibitem{Buch_Harvey_Giorgino_Anderson_DeFabritiis_2010}
I.~Buch, M.~J. Harvey, T.~Giorgino, D.~P. Anderson, and G.~De~Fabritiis.
\newblock High-throughput all-atom molecular dynamics simulations using
  distributed computing.
\newblock {\em Journal of Chemical Information and Modeling}, 50(3):397--403,
  Mar 2010.

\bibitem{BuchFabritiis_PNAS11_Binding}
Ignasi Buch, Toni Giorgino, and Gianni De~Fabritiis.
\newblock {Complete reconstruction of an enzyme-inhibitor binding process by
  molecular dynamics simulations}.
\newblock {\em Proceedings of the National Academy of Sciences},
  108(25):10184--10189, June 2011.

\bibitem{BucheteHummer_JPCB08}
Nicaolae~V. Buchete and Gerhard Hummer.
\newblock {Coarse Master Equations for Peptide Folding Dynamics}.
\newblock {\em J. Phys. Chem. B}, 112:6057--6069, 2008.

\bibitem{Buchner_BBA11_ProteinFoldingKinetics}
Ginka~S. Buchner, Ronan~D. Murphy, Nicolae-Viorel Buchete, and Jan Kubelka.
\newblock Dynamics of protein folding: Probing the kinetic network of
  folding--unfolding transitions with experiment and theory.
\newblock {\em Biochimica et Biophysica Acta}, 1814:1001--1020, 2011.

\bibitem{ChoderaEtAl_JCP07}
J.~D. Chodera, K.~A. Dill, N.~Singhal, V.~S. Pande, W.~C. Swope, and J.~W.
  Pitera.
\newblock {Automatic discovery of metastable states for the construction of
  Markov models of macromolecular conformational dynamics}.
\newblock {\em J. Chem. Phys.}, 126:155101, 2007.

\bibitem{ChungLuoisEaton_Science12_TransitionPathTimes}
H.S. Chung, J.M. Louis, and W.A. Eaton.
\newblock Single-molecule fluorescence experiments determine protein folding
  transition path times.
\newblock {\em Science}, 335:981--984, 2012.

\bibitem{CoifmanLafon_PNAS05_DiffusionMaps}
R.~R. Coifman, S.~Lafon, A.~B. Lee, M.~Maggioni, B.~Nadler, F.~Warner, and
  S.~W. Zucker.
\newblock Geometric diffusions as a tool for harmonic analysis and structure
  definition of data: Diffusion maps.
\newblock {\em Proc. Natl. Acad. Sci. USA}, 102:7426--7431, 2005.

\bibitem{Daidone_PlosOne10_MR121Kinetics}
Isabella Daidone, Hannes Neuweiler, S\"{o}ren Doose, Markus Sauer, and
  Jeremy~C. Smith.
\newblock {Hydrogen-bond driven loop-closure kinetics in unfolded polypeptide
  chains}.
\newblock {\em PloS One}, 6:e1000645+, 2010.

\bibitem{DasguptaLong_JComputSystSci05_kCenters}
S.~Dasgupta and P.~Long.
\newblock {Performance guarantees for hierarchical clustering}.
\newblock {\em J. Comput. Syst. Sci.}, 70(4):555--569, June 2005.

\bibitem{deGrootDauraMarkGrubmuller_JMB301_299}
B.~de~Groot, X.~Daura, A.~Mark, and H.~Grubm\"{u}ller.
\newblock {Essential Dynamics of Reversible Peptide Folding: Memory-free
  Conformational Dynamics Governed by Internal Hydrogen Bonds}.
\newblock {\em J. Mol. Bio.}, 301:299--313, 2001.

\bibitem{DeuflhardWeber_PCCA}
P.~Deuflhard and M.~Weber.
\newblock {Robust Perron cluster analysis in conformation dynamics}.
\newblock {\em ZIB Report}, 03-09, 2003.

\bibitem{DjurdjevacSarichSchuette_MMS10_EigenvalueError}
Natasa Djurdjevac, Marco Sarich, and Christof Sch\"{u}tte.
\newblock {Estimating the eigenvalue error of Markov State Models}.
\newblock {\em Multiscale Model. Simul.}, 10:61--81, 2012.

\bibitem{EisenmesserKayKern_Nature2005}
Elan~Z. Eisenmesser, Oscar Millet, Wladimir Labeikovsky, Dmitry~M. Korzhnev,
  Magnus Wolf-Watz, Daryl~A. Bosco, Jack~J. Skalicky, Lewis~E. Kay, and
  Dorothee Kern.
\newblock {Intrinsic dynamics of an enzyme underlies catalysis}.
\newblock {\em Nature}, 438(7064):117--121, November 2005.

\bibitem{feenstra_improving_1999}
K.~Anton Feenstra, Berk Hess, and Herman J.~C. Berendsen.
\newblock Improving efficiency of large time-scale molecular dynamics
  simulations of hydrogen-rich systems.
\newblock {\em Journal of Computational Chemistry}, 20(8):786--798, 1999.

\bibitem{FischerSmith_PNAS102_6873}
S.~Fischer, B.~Windshuegel, D.~Horak, K.~C. Holmes, and J.~C. Smith.
\newblock {Structural mechanism of the recovery stroke in the Myosin molecular
  motor}.
\newblock {\em Proc. Natl. Acad. Sci. USA}, 102:6873--6878, 2005.

\bibitem{GansenSeidel_PNAS2009_Nucleosome}
Alexander Gansen, Alessandro Valeri, Florian Hauger, Suren Felekyan, Stanislav
  Kalinin, Katalin T\'{o}th, J\"{o}rg Langowski, and Claus A.~M. Seidel.
\newblock {Nucleosome disassembly intermediates characterized by
  single-molecule FRET}.
\newblock {\em Proc. Natl. Acad. Sci. USA}, 106(36):15308--15313, September
  2009.

\bibitem{GebhardRief_PNAS10_AFMEnergyLandscapeProtein}
Gebhardt, Thomas Bornschl\"{o}gl, and Matthias Rief.
\newblock {Full distance-resolved folding energy landscape of one single
  protein molecule}.
\newblock {\em Proc. Natl. Acad. Sci. USA}, 107(5):2013--2018, February 2010.

\bibitem{Harvey_Giupponi_Fabritiis_2009}
M.~J. Harvey, G.~Giupponi, and G.~De Fabritiis.
\newblock Acemd: Accelerating biomolecular dynamics in the microsecond time
  scale.
\newblock {\em Journal of Chemical Theory and Computation}, 5(6):1632--1639,
  Jun 2009.

\bibitem{HeldEtAl_BiophysJ10_AssociationTPT}
Martin Held, Philipp Metzner, and Frank No{\'e}.
\newblock Mechanisms of protein-ligand association and its modulation by
  protein mutations.
\newblock {\em Biophys. J.}, 100:701--710, 2011.

\bibitem{Hinrichs_Pande_2007}
Nina~Singhal Hinrichs and Vijay~S. Pande.
\newblock Calculation of the distribution of eigenvalues and eigenvectors in
  markovian state models for molecular dynamics.
\newblock {\em The Journal of Chemical Physics}, 126(24):244101--244101--11,
  Jun 2007.

\bibitem{HuangCaflisch_PlosCB11_SmallMoleculeUnbinding}
Danzhi Huang and Amedeo Caflisch.
\newblock {The Free Energy Landscape of Small Molecule Unbinding}.
\newblock {\em PLoS Comput Biol}, 7(2):e1002002+, February 2011.

\bibitem{HubnerShakhnovich_PNAS96_EnsembleFolding}
Isaac~A. Hubner, Eric~J. Deeds, and Eugene~I. Shakhnovich.
\newblock {Understanding ensemble protein folding at atomic detail}.
\newblock {\em Proc. Natl. Acad. Sci. USA}, 103(47):17747--17752, November
  2006.

\bibitem{JaegerEtAl_PNAS06}
M.~J{\"{a}}ger, Y.~Zhang, J.~Bieschke, H.~Nguyen, M.~Dendle, M.~E. Bowman,
  J.~P. Noel, M.~Gruebele, and J.~W. Kelly.
\newblock {Structure-function-folding relationship in a WW domain}.
\newblock {\em Proc. Natl. Acad. Sci. USA}, 103:10648--10653, 2006.

\bibitem{Jain::10}
Abhinav Jain, Rainer Hegger, and Gerhard Stock.
\newblock Hidden complexity of protein free-energy landscapes revealed by
  principal component analysis by parts.
\newblock {\em The Journal of Physical Chemistry Letters}, 1(19):2769--2773,
  2010.

\bibitem{Jolliffe_PCA-book}
I.T. Jolliffe.
\newblock {\em Principal Component Analysis}.
\newblock Springer, New York, 2 edition, 2002.

\bibitem{jorgensen_comparison_1983}
William~L. Jorgensen, Jayaraman Chandrasekhar, Jeffry~D. Madura, Roger~W.
  Impey, and Michael~L. Klein.
\newblock Comparison of simple potential functions for simulating liquid water.
\newblock {\em The Journal of Chemical Physics}, 79(2):926--935, July 1983.

\bibitem{Joung_Cheatham_2008}
In~Suk Joung and Thomas~E. Cheatham.
\newblock Determination of alkali and halide monovalent ion parameters for use
  in explicitly solvated biomolecular simulations.
\newblock {\em The Journal of Physical Chemistry. B}, 112(30):9020--9041, Jul
  2008.
\newblock PMID: 18593145 PMCID: 2652252.

\bibitem{Kabsch::76}
W.~Kabsch.
\newblock A solution for the best rotation to relate two sets of vectors.
\newblock {\em Acta Crystallographica Section A}, 32(5):922--923, Sep 1976.

\bibitem{KarpenBrooks_Biochemistry93_Clustering}
Mary~E. Karpen, Douglas~J. Tobias, and Charles~L. Brooks.
\newblock {Statistical clustering techniques for the analysis of long molecular
  dynamics trajectories: analysis of 2.2-ns trajectories of YPGDV}.
\newblock {\em Biochemistry}, 32(2):412--420, January 1993.

\bibitem{Kasper_2006}
Lawryn~H Kasper, Tomofusa Fukuyama, Michelle~A Biesen, Fay\c{c}al Boussouar,
  Caili Tong, Antoine de~Pauw, Peter~J Murray, Jan M~A van Deursen, and Paul~K
  Brindle.
\newblock Conditional knockout mice reveal distinct functions for the global
  transcriptional coactivators {CBP} and p300 in {T}-cell development.
\newblock {\em Molecular and cellular biology}, 26(3):789--809, Feb 2006.
\newblock PMID: 16428436.

\bibitem{KaufmanRousseeuw_kMedoids}
L.~Kaufman and P.J. Rousseeuw.
\newblock {\em Statistical Data Analysis Based on the L\_1-Norm and Related
  Methods}, chapter Clustering by means of Medoids, pages 405--416.
\newblock North-Holland, 1987.

\bibitem{KellerPrinzNoe_ChemPhysReview11}
Bettina Keller, Jan-Hendrik Prinz, and Frank No{\'e}.
\newblock Markov models and dynamical fingerprints: Unraveling the complexity
  of molecular kinetics.
\newblock {\em Chem. Phys.}, 396:92--107, 2012.

\bibitem{KobitskiNienhaus_NucleicAcidsRes07}
Andrei~Y. Kobitski, Alexander Nierth, Mark Helm, Andres J\"{a}schke, and
  G.~Ulrich Nienhaus.
\newblock {Mg2+ dependent folding of a Diels-Alderase ribozyme probed by
  single-molecule FRET analysis}.
\newblock {\em Nucleic Acids Res.}, 35:2047--2059, 2007.

\bibitem{KrivovKarplus_PNAS101_14766}
S.~V. Krivov and M.~Karplus.
\newblock {Hidden complexity of free energy surfaces for peptide (protein)
  folding}.
\newblock {\em Proc. Nat. Acad. Sci. USA}, 101:14766--14770, 2004.

\bibitem{KubeWeber_JCP07_CoarseGraining}
Susanna Kube and Marcus Weber.
\newblock {A coarse graining method for the identification of transition rates
  between molecular conformations}.
\newblock {\em J. Chem. Phys.}, 126(2):024103+, 2007.

\bibitem{Tong::90}
{L. Tong}, {V.C. Soon}, {Y.F. Huang}, and {R. Liu}.
\newblock A{MUSE}: a new blind identification algorithm.
\newblock {\em Circuits and Systems}, 3:1784--1787, 1990.

\bibitem{LindorffLarsenEtAl_Science11_AntonFolding}
Kresten Lindorff-Larsen, Stefano Piana, Ron~O. Dror, and David~E. Shaw.
\newblock How fast-folding proteins fold.
\newblock {\em Science}, 334:517--520, 2011.

\bibitem{Lindorff_2010}
Kresten Lindorff-Larsen, Stefano Piana, Kim Palmo, Paul Maragakis, John~L
  Klepeis, Ron~O Dror, and David~E Shaw.
\newblock Improved side-chain torsion potentials for the {Amber ff99SB} protein
  force field.
\newblock {\em Proteins}, 78(8):1950--1958, Jun 2010.
\newblock PMID: 20408171 PMCID: PMC2970904.

\bibitem{Lloyd_IEEE82_kMeans}
S.P. Lloyd.
\newblock Least squares quantization in pcm.
\newblock {\em IEEE Transactions on Information Theory}, 28:129--137, 1982.

\bibitem{MetznerSchuetteVandenEijnden_TPT}
P.~Metzner, C.~Sch\"{u}tte, and E.~Vanden Eijnden.
\newblock {Transition Path Theory for Markov Jump Processes}.
\newblock {\em Multiscale Model. Simul.}, 7:1192--1219, 2009.

\bibitem{Xie_PRL05_PowerLaw}
W.~Min, G.~Luo, B.~J. Cherayil, S.~C. Kou, and X.~S. Xie.
\newblock {Observation of a Power-Law Memory Kernel for Fluctuations within a
  Single Protein Molecule}.
\newblock {\em Phys. Rev. Lett.}, 94:198302+, 2005.

\bibitem{Mitsutake::11}
Ayori Mitsutake, Hiromitsu Iijima, and Hiroshi Takano.
\newblock Relaxation mode analysis of a peptide system: Comparison with
  principal component analysis.
\newblock {\em The Journal of Chemical Physics}, 135(16):164102, 2011.

\bibitem{Mohan_Uversky_2006}
Amrita Mohan, Christopher~J. Oldfield, Predrag Radivojac, Vladimir Vacic,
  Marc~S. Cortese, A.~Keith Dunker, and Vladimir~N. Uversky.
\newblock Analysis of molecular recognition features ({MoRFs}).
\newblock {\em Journal of Molecular Biology}, 362(5):1043--1059, Oct 2006.

\bibitem{Molgedey::94}
L.~Molgedey and H.~G. Schuster.
\newblock Separation of a mixture of independent signals using time delayed
  correlations.
\newblock {\em Phys. Rev. Lett.}, 72:3634--3637, Jun 1994.

\bibitem{MuffCaflisch_Proteins07}
Stefanie Muff and Amedeo Caflisch.
\newblock {Kinetic analysis of molecular dynamics simulations reveals changes
  in the denatured state and switch of folding pathways upon single-point
  mutation of a -sheet miniprotein}.
\newblock {\em Proteins}, 70:1185--1195, 2007.

\bibitem{Naritomi::11}
Yusuke Naritomi and Sotaro Fuchigami.
\newblock Slow dynamics in protein fluctuations revealed by time-structure
  based independent component analysis: The case of domain motions.
\newblock {\em The Journal of Chemical Physics}, 134(6):065101, 2011.

\bibitem{NeubauerSeidel_JACS2007_Rhodamine}
Heike Neubauer, Natalia Gaiko, Sylvia Berger, J\"{o}rg Schaffer, Christian
  Eggeling, Jennifer Tuma, Laurent Verdier, Claus~A. Seidel, Christian
  Griesinger, and Andreas Volkmer.
\newblock {Orientational and dynamical heterogeneity of rhodamine 6G terminally
  attached to a DNA helix revealed by NMR and single-molecule fluorescence
  spectroscopy.}
\newblock {\em J. Am. Chem. Soc.}, 129(42):12746--12755, October 2007.

\bibitem{NeuweilerEtAl_JMB07}
Hannes Neuweiler, Marc L{\"{o}}llmann, S{\"{o}}ren Doose, and M.~Sauer.
\newblock {Dynamics of Unfolded Polypeptide Chains in Crowded Environment
  Studied by Fluorescence Correlation Spectroscopy}.
\newblock {\em J. Mol. Biol.}, 365:856--869, 2007.

\bibitem{NoeDaidoneSmithAmadei_JPCB08_QGE}
F.~No\'{e}, I.~Daidone, J.~C. Smith, A.~di~Nola, and A.~Amadei.
\newblock {Solvent Electrostriction Driven Peptide Folding revealed by
  Quasi-Gaussian Entropy Theory and Molecular Dynamics Simulation}.
\newblock {\em Journal of Physical Chemistry B}, 112:11155--11163, 2008.

\bibitem{NoeFischer_CurrOpin08_TransitionNetworks}
F.~No\'{e} and S.~Fischer.
\newblock {Transition networks for modeling the kinetics of conformational
  transitions in macromolecules}.
\newblock {\em Curr. Opin. Struc. Biol.}, 18:154--162, 2008.

\bibitem{NoeHorenkeSchutteSmith_JCP07_Metastability}
F.~No\'{e}, I.~Horenko, C.~Sch\"{u}tte, and J.~C. Smith.
\newblock {Hierarchical Analysis of Conformational Dynamics in Biomolecules:
  Transition Networks of Metastable States}.
\newblock {\em J. Chem. Phys.}, 126:155102, 2007.

\bibitem{NoeKrachtusSmithFischer_JCTC06_TransitionNetworks}
F.~No\'{e}, D.~Krachtus, J.~C. Smith, and S.~Fischer.
\newblock {Transition Networks for the Comprehensive Characterization of
  Complex Conformational Change in Proteins}.
\newblock {\em J. Chem. Theo. Comp.}, 2:840--857, 2006.

\bibitem{Noe_JCP08_TSampling}
Frank No{\'e}.
\newblock {Probability Distributions of Molecular Observables computed from
  Markov Models}.
\newblock {\em J. Chem. Phys.}, 128:244103, 2008.

\bibitem{NoeEtAl_PNAS11_Fingerprints}
Frank No{\'e}, S{\"o}ren Doose, Isabella Daidone, Marc L{\"o}llmann, John~D.
  Chodera, Markus Sauer, and Jeremy~C. Smith.
\newblock Dynamical fingerprints for probing individual relaxation processes in
  biomolecular dynamics with simulations and kinetic experiments.
\newblock {\em Proc. Natl. Acad. Sci. USA}, 108:4822--4827, 2011.

\bibitem{Noe_MMS12_VariationalPrinciple}
Frank No{\'e} and Feliks N{\"u}ske.
\newblock A variational approach to modeling slow processes in stochastic
  dynamical systems.
\newblock {\em Multiscale Model. Simul. (submitted, available at:
  http://publications.mi.fu-berlin.de/1109/)}, 2012.

\bibitem{NoeSchuetteReichWeikl_PNAS09_TPT}
Frank No{\'e}, Christof Sch{\"u}tte, Eric Vanden-Eijnden, Lothar Reich, and
  Thomas~R. Weikl.
\newblock Constructing the full ensemble of folding pathways from short
  off-equilibrium simulations.
\newblock {\em Proc. Natl. Acad. Sci. USA}, 106:19011--19016, 2009.

\bibitem{Nienhaus_Nature00}
Andreas Ostermann, Robert Waschipky, Fritz~G. Parak, and Ulrich~G. Nienhaus.
\newblock {Ligand binding and conformational motions in myoglobin}.
\newblock {\em Nature}, 404:205--208, 2000.

\bibitem{PanRoux_JCP08_MarkovModelPath}
Albert~C. Pan and Beno\^{i}t Roux.
\newblock {Building Markov state models along pathways to determine free
  energies and rates of transitions}.
\newblock {\em J. Chem. Phys.}, 129(6):064107+, August 2008.

\bibitem{PirchiHaran_NatureComms11_SingleMoleculeFRET}
Menahem Pirchi, Guy Ziv, Inbal Riven, Sharona~Sedghani Cohen, Nir Zohar, Yoav
  Barak, and Gilad Haran.
\newblock Single-molecule fluorescence spectroscopy maps the folding landscape
  of a large protein.
\newblock {\em Nature Comms}, 2:493, 2011.

\bibitem{PrinzChoderaNoe_PRL11_RateTheory}
Jan-Hendrik Prinz, John~D. Chodera, and Frank No{\'e}.
\newblock Robust rate estimate from spectral estimation.
\newblock {\em Phys. Rev. Lett. (submitted)}, 2011.

\bibitem{PrinzEtAl_JCP10_MSM1}
Jan-Hendrik Prinz, Hao Wu, Marco Sarich, Bettina Keller, Martin Fischbach,
  Martin Held, John~D. Chodera, Christof Sch{\"u}tte, and Frank No{\'e}.
\newblock Markov models of molecular kinetics: Generation and validation.
\newblock {\em J. Chem. Phys.}, 134:174105, 2011.

\bibitem{RadhakrishnanWrite_Cell97_KIDKIX}
Ishwar Radhakrishnan, Gabriela~C P{\'e}rez-Alvarado, David Parker, H.Jane
  Dyson, Marc~R Montminy, and Peter~E Wright.
\newblock Solution structure of the kix domain of cbp bound to the
  transactivation domain of creb: A model for activator:coactivator
  interactions.
\newblock {\em Cell}, 91:741--752, 1997.

\bibitem{RaoCaflisch_JMB342_299}
F.~Rao and A.~Caflisch.
\newblock {The Protein Folding Network}.
\newblock {\em J. Mol. Bio.}, 342:299--306, 2004.

\bibitem{Ritz_JReineAngewMathe09_Variationsprobleme}
Walter Ritz.
\newblock {\"U}ber eine neue methode zur l{\"o}sung gewisser variationsprobleme
  der mathematischen physik.
\newblock {\em Journal f{\"u}r die Reine und Angewandte Mathematik}, 135:1--61,
  1909.

\bibitem{RohrdanzClementi_JCP134_DiffMaps}
M.~A. Rohrdanz, W.~Zheng, M.~Maggioni, and C.~Clementi.
\newblock Determination of reaction coordinates via locally scaled diffusion
  map.
\newblock {\em J. Chem. Phys.}, 134(124116), 2011.

\bibitem{SadiqNoeFabritiis_PNAS12_HIV}
S.~K. Sadiq, F.~No{\'e}, and G.~de~Fabritiis.
\newblock Kinetic characterization of the critical step in hiv-1 protease
  maturation.
\newblock {\em Proc. Natl. Acad. Sci. USA (in press)}, 2012.

\bibitem{Santoso_PNAS2009_FretPolymerase}
Yusdi Santoso, Catherine~M. Joyce, Olga Potapova, Ludovic Le~Reste, Johannes
  Hohlbein, Joseph~P. Torella, Nigel D.~F. Grindley, and Achillefs~N.
  Kapanidis.
\newblock {Conformational transitions in DNA polymerase I revealed by
  single-molecule FRET}.
\newblock {\em Proc. Natl. Acad. Sci. USA}, 107(2):715--720, January 2010.

\bibitem{SarichNoeSchuette_MMS09_MSMerror}
Marco Sarich, Frank No{\'e}, and Christof Sch{\"u}tte.
\newblock On the approximation error of markov state models.
\newblock {\em SIAM Multiscale Model. Simul.}, 8:1154--1177, 2010.

\bibitem{SchultheisHirschbergerCarstensTavan_JCTC1_515}
V.~Schultheis, T.~Hirschberger, H.~Carstens, and P.~Tavan.
\newblock {Extracting Markov Models of Peptide Conformational Dynamics from
  Simulation Data}.
\newblock {\em J. Chem. Theory Comp.}, 1:515--526, 2005.

\bibitem{SchuetteFischerHuisingaDeuflhard_JCompPhys151_146}
C.~Sch\"{u}tte, A.~Fischer, W.~Huisinga, and P.~Deuflhard.
\newblock {A Direct Approach to Conformational Dynamics based on Hybrid Monte
  Carlo}.
\newblock {\em J. Comput. Phys.}, 151:146--168, 1999.

\bibitem{SeeberCaflisch_JCC11_Wordom}
M.~Seeber, A.~Felline, F.~Raimondi, S.~Muff, Ran Friedman, F.~Rao, A.~Caflisch,
  and F.~Fanelli.
\newblock Wordom: a user-friendly program for the analysis of molecular
  structures, trajectories, and free energy surfaces.
\newblock {\em J. Comp. Chem.}, 6:1183--1194, 2011.

\bibitem{SenneSchuetteNoe_JCTC12_EMMA1.2}
M.~Senne, B.~Trendelkamp-Schroer, A.S.J.S. Mey, C.~Sch{\"u}tte, and F.~No{\'e}.
\newblock Emma - a software package for markov model building and analysis.
\newblock {\em J. Chem. Theory and Comput.}, 8:2223--2238, 2012.

\bibitem{Shaw_Science10_Anton}
David~E. Shaw, Paul Maragakis, Kresten Lindorff-Larsen, Stefano Piana, Ron~O.
  Dror, Michael~P. Eastwood, Joseph~A. Bank, John~M. Jumper, John~K. Salmon,
  Yibing Shan, and Willy Wriggers.
\newblock {Atomic-Level Characterization of the Structural Dynamics of
  Proteins}.
\newblock {\em Science}, 330(6002):341--346, October 2010.

\bibitem{Shoemaker_Portman_Wolynes_2000}
Benjamin~A. Shoemaker, John~J. Portman, and Peter~G. Wolynes.
\newblock Speeding molecular recognition by using the folding funnel: The
  fly-casting mechanism.
\newblock {\em Proceedings of the National Academy of Sciences},
  97(16):8868--8873, Aug 2000.

\bibitem{SinghalPande_JCP123_204909}
N.~Singhal and V.~S. Pande.
\newblock {Error analysis and efficient sampling in Markovian state models for
  molecular dynamics}.
\newblock {\em J. Chem. Phys.}, 123:204909, 2005.

\bibitem{SriramanKevrekidisHummer_JPCB109_6479}
S.~Sriraman, I.~G. Kevrekidis, and G.~Hummer.
\newblock {Coarse Master Equation from Bayesian Analysis of Replica Molecular
  Dynamics Simulations}.
\newblock {\em J. Phys. Chem. B}, 109:6479--6484, 2005.

\bibitem{StiglerRief_Science11_CalmodulinFoldingNetwork}
Johannes Stigler, Fabian Ziegler, Anja Gieseke, J.~Christof~M. Gebhardt, and
  Matthias Rief.
\newblock The complex folding network of single calmodulin molecules.
\newblock {\em Science}, 334:512--516, 2011.

\bibitem{Sugase_Dyson_Wright_2007}
Kenji Sugase, H.~Jane Dyson, and Peter~E. Wright.
\newblock Mechanism of coupled folding and binding of an intrinsically
  disordered protein.
\newblock {\em Nature}, 447(7147):1021--1025, Jul 2007.

\bibitem{SwopePiteraSuits_JPCB108_6571}
W.~C. Swope, J.~W. Pitera, and F.~Suits.
\newblock {Describing protein folding kinetics by molecular dynamics
  simulations: 1. Theory}.
\newblock {\em J. Phys. Chem. B}, 108:6571--6581, 2004.

\bibitem{Uversky_2011}
Vladimir~N. Uversky.
\newblock Intrinsically disordered proteins from {A to Z}.
\newblock {\em The International Journal of Biochemistry \& Cell Biology},
  43(8):1090--1103, Aug 2011.

\bibitem{VoelzPande_JACS10_NTL9}
Vincent~A. Voelz, Gregory~R. Bowman, Kyle Beauchamp, and Vijay~S. Pande.
\newblock {Molecular Simulation of ab Initio Protein Folding for a Millisecond
  Folder NTL9}.
\newblock {\em J. Am. Chem. Soc.}, 132(5):1526--1528, February 2010.

\bibitem{Wales}
D.~J. Wales.
\newblock {\em {Energy Landscapes}}.
\newblock Cambridge University Press, Cambridge, 2003.

\bibitem{Weber_ImprovedPCCA}
M.~Weber.
\newblock {Improved Perron cluster analysis}.
\newblock {\em ZIB Report}, 03-04, 2003.

\bibitem{WensleyClark_Nature09_FrustrationHelix}
Beth~G. Wensley, Sarah Batey, Fleur A.~C. Bone, Zheng~M. Chan, Nuala~R.
  Tumelty, Annette Steward, Lee~G. Kwa, Alessandro Borgia, and Jane Clarke.
\newblock {Experimental evidence for a frustrated energy landscape in a
  three-helix-bundle protein family}.
\newblock {\em Nature}, 463(7281):685--688, February 2010.

\end{thebibliography}

\end{document}